\documentclass[11pt]{article}
\usepackage{geometry}                
\usepackage{graphicx}
\usepackage{amssymb}
\usepackage{amsmath}
\usepackage{amsfonts}
\usepackage{graphicx}
\usepackage{epstopdf}
\usepackage{color}
\usepackage{cite}

\newcommand{\be}{\begin{equation}}
\newcommand{\ee}{\end{equation}}
\newcommand{\bea}{\begin{eqnarray}}
\newcommand{\eea}{\end{eqnarray}}
\newcommand{\beas}{\begin{eqnarray*}}
\newcommand{\eeas}{\end{eqnarray*}}
\newcommand{\avg}[1]{\left\langle{#1}\right\rangle}

\usepackage{float}
\usepackage[titletoc,title]{appendix}

\usepackage{parskip}

\title{Portfolio Optimization under Expected Shortfall: Contour Maps of Estimation Error}
\date{}
\author{Fabio Caccioli$^{1,2}$, Imre Kondor$^{3,4}$ and G\'abor Papp$^5$\\
{\it 1- University College London, Department of Computer Science,} \\
{\it London, WC1E 6BT, UK} \\
{\it 2- Systemic Risk Centre, London School of Economics and Political Sciences, London, UK}\\
{\it 3-Parmenides Foundation, Pullach, Germany}\\
{\it 4- Department of Investment and Corporate Finance, Corvinus University of Budapest,} \\
{\it Budapest, Hungary}\\
{\it 5- E\"otv\"os Lor\'and University, Institute for Physics, Budapest, Hungary }
}
\begin{document}
\bibliographystyle{unsrt}

\maketitle

\begin{abstract} 
The contour maps of the error of historical resp. parametric estimates for large random portfolios optimized under the risk measure Expected Shortfall (ES) are constructed. Similar maps for the sensitivity of the portfolio weights to small changes in the returns as well as the VaR of the ES-optimized portfolio are also presented, along with results for the distribution of portfolio weights over the random samples and for the out-of-sample and in-the-sample estimates for ES. The contour maps allow one to quantitatively determine the sample size (the length of the time series) required by the optimization for a given number of different assets in the portfolio, at a given confidence level and a given level of relative estimation error. The necessary sample sizes invariably turn out to be unrealistically large for any reasonable choice of the number of assets and the confidence level. These results are obtained via analytical calculations based on methods borrowed from the statistical physics of random systems, supported by numerical simulations.
\end{abstract}

\section{Introduction}
Risk measurement and portfolio optimization are two complementary aspects of portfolio theory. Both assume that  the future will statistically resemble the past, but while risk measurement is trying to forecast the risk of an existing portfolio, optimization is attempting to choose the composition of the portfolio in such a way as to minimize risk at a given level of  expected return (or maximize return at a given level of risk). In the case of large institutional portfolios both tasks require a large number of input data, i.e. large samples of observed returns. The inherent difficulty both of risk measurement and optimization lies in the fact that these large sample sizes are typically hard if not impossible to achieve in practice. On the portfolio selection side in particular, the difficulty is further aggravated by the fact that here the sample size must be large not only compared to one, but also to the size (measured in the number of different assets) of the portfolio. In order to have samples exceeding the dimensions of large institutional portfolios with numbers of assets in the hundreds or thousands, one would need either a high sampling frequency, or a long look-back period, and preferably both. The length of the look-back period is limited by considerations of lack of stationarity: some of the assets in the portfolio may be shares of firms that have not been in existence for a very long time, the economic and monetary environment may have changed, new regulatory constraints may have been introduced, etc. As for the sampling frequency, it is limited by the frequency at which portfolios can practically be rebalanced, so we assume in the following that the portfolio manager uses such low frequency (weekly or rather monthly) data. This means that the sample size (the length of the time series $T$) is always limited, while the dimension $N$ of institutional portfolios (the number of different assets) is typically very large, and the condition $N/T \ll 1$ necessary for reliable and stable estimates is almost always violated in practice. In this paper we will consider situations where $N$ and $T$ are both very large, but their ratio is finite.

Both risk measurement and optimization require a metric, a measure of risk. In Markowitz's original theory of portfolio optimization \cite{markowitz1952Portfolio} the risk measure was chosen to be the volatility of the return data, identified with the variance of the observed time series. If one measures risk in terms of the variance, this equally penalizes large negative as well as positive fluctuations. The symmetric treatment of loss and gain was considered unjustified from the point of view of the investor, therefore the idea of downside risk measures, focusing solely on losses, was introduced very early, already by Markowitz \cite{markowitz1959Portfolio}, in the form of the semivariance. A few decades later, in the aftermath of the meltdown on Black Monday in October 1987, and the collapse of the US savings and loan industry in the late 80's and early 90's, it was realized that the really lethal danger lurked in the far tail of the loss distribution and that the probability of such catastrophic events was much higher than what could be estimated on the basis of the normal distribution. Value at Risk (VaR) started to sporadically appear around the end of the 80's as an attempt at grasping this tail risk. It was adopted as the metric in the daily risk reports at J.P. Morgan, later widely spread by their RiskMetrics methodology \cite{morgan1995riskmetrics} that for a certain period became a sort of industry standard. VaR's status was further elevated when it became adopted as the ``official'' risk measure by international regulation in 1996 \cite{basle1996Overview}.

Value at Risk (VaR) is a high quantile of the profit and loss distribution, the threshold the portfolio's loss will not exceed with probability $\alpha$. In practice, the typical values of this confidence level were chosen to be 0.90, 0.95, or 0.99.

In spite of its undeniable merits, VaR came under criticism very soon for its lack of subadditivity, which violated the principle of diversification, and also for its failure to say anything about the behaviour of the distribution beyond the VaR quantile. By an axiomatic approach to the problem of risk measures, Artzner et al. \cite{Artzner1999Coherent} introduced the concept of coherent measures that are free of these shortcomings by construction. The simplest representative of coherent  measures is Expected Shortfall (ES), the average risk above a high quantile that can be chosen to be equal to the VaR threshold. For this reason ES is also called conditional VaR or CVaR. 

As a conditional average, ES is sensitive not only to the total mass of fluctuations above the quantile, but also to their distribution. This, and its coherence proved by Pflug \cite{pflug2000some} and Acerbi and Tasche \cite{acerbi2002Expected,acerbi2002On} has made it popular among theoreticians, but increasingly also among practitioners. Recently, it has also been embraced by regulation \cite{basel2013Fundamental,basel2014Fundamental} that envisages a confidence level of 0.975 for ES.

Today VaR and ES are the two most frequently used risk measures. It is therefore important to investigate their statistical properties, especially in the typical high-dimensional setting. The lack of sufficient data which, for large portfolios, can be very serious for any risk measure is particularly grave in the case of downside risk measures (such as VaR and ES) that discard most of the observed data except those above the high quantile.

A comprehensive recent treatment of the risk measurement aspect of the problem is due to Danielsson and Zhou \cite{danielsson2015Why}. Our purpose here is to look into the complementary problem: that of portfolio selection. If we knew the true probability distribution of returns, it would be straightforward to determine the optimal composition of the portfolio (the optimal portfolio weights) and calculate the true value of Expected Shortfall. The true distribution of returns is, however, unknown. What we may have in practice is only a finite sample, and the optimal weights and ES have to be estimated on the basis of this information. The resulting weights and ES will deviate from their ``true'' values (that would be obtained in an infinitely large stationary sample), and the deviation can be expected to be the stronger the shorter the length $T$ of the sample and the larger the dimension $N$ of the portfolio. In addition, in a different sample we would obtain a different estimate: there is a distribution of ES and of the optimal weights over the samples.

How can one cope with the estimation error arising from this relative scarcity of data? In actual practice, where one really has to live with a single sample of limited size, one may use cross-validation or bootstrap \cite{hastie2008Elements}. In the present theoretical work, we choose an alternative procedure  to mimic historical estimation: Instead of the unknown underlying process, we consider a simple, easily manageable one, such as a multivariate Gaussian process, where the true ES is easy to obtain, thereby creating a firm basis for comparison. Then we calculate ES for a large number of random samples of length $T$ drawn from this underlying distribution,  average ES over these random samples and finally compare this average to its true value. This excercise will give us an idea about how large the estimation error can be for a given dimension $N$ and sample size $T$, and we may expect that the optimization of portfolios under ES with a non-stationary and fat tailed real-life process will suffer from an even more serious estimation error than its Gaussian counterpart. In other words, we expect that the estimation error for a stationary Gaussian underlying process is a lower bound for the estimation error for real-life processes.

This program can certainly be carried out by numerical simulations. To obtain analytical results for ES optimization is, however, nontrivial and we are not aware of any analytical approach using standard methods of probability theory or statistics that could be applied to this problem. However, methods borrowed from the theory of random systems, in particular the replica method \cite{mezard1987Spin}, do offer the necessary tools in the special case of a Gaussian underlying process, and these are the methods we are going to apply here. For the sake of simplicity, we will also assume that the returns are independent, identically distributed normal variables, although the assumptions of independence and identical distribution could be relaxed and the calculation could still be carried through without essentially changing the conclusions. We will briefly discuss the case of normal variables with an arbitrary (but invertible) covariance matrix later in the paper.

The Gaussian assumption is more serious: if we drop it, we are no longer able to perform the calculations analytically. Numerical simulations remain feasible, however, and we will carry out simulations for the case of independent Student-distributed returns (with $\nu=3$ degrees of freedom, asymptotically falling off like $x^{-4}$), in order to see how much difference the fat tailed character of this distribution makes in the estimation error. (We will also consider a Student distribution with $\nu=10$ to show how the numerical results approach the Gaussian case.) As expected, the large fluctuations at the tail result in a deterioration of the estimated ES. This supports our conjecture that the estimation error found in the case of normally distributed returns is a lower bound to the estimation error for other, more realistic distributions, and for that reason the present exercise has a message for portfolio optimization in general.

The analytical technique we are going to apply enables us to calculate the relative error of ES and the distribution of optimal portfolio weights averaged over the random Gaussian samples, but does not provide information (at least not without a great deal of additional effort) about how strongly these quantities fluctuate from sample to sample. In the limit of large portfolio sizes the distribution of estimated Expected Shortfall and its error can be expected to become sharp, with ES and the estimation error becoming independent of the samples\footnote{For a rigorous mathematical proof of this ``self-averaging'' in the case of related models of statistical physics see \cite{guerra2003TheInfinite,guerra2002TheThermodynamic}}. In order to acquire  information about the distribution of these estimates over the samples, we will resort to numerical simulations again. We find that the distribution of the estimated ES over the samples is becoming reasonably sharp already for $N$'s in the range of a few hundred, so the sample average can give us a good idea about the estimation error. In the vicinity of some special, critical points, however, the average estimation error can grow beyond any bound, and there its  fluctuations also diverge.

As we have already mentioned, the lack of sufficient information leads to large errors in the estimation of optimal weights and overall portfolio risk under any risk measure. In the case of e.g. the variance as risk measure, it is well known that the sample size $T$ must be much larger than the dimension $N$ of the portfolio, if we wish to obtain a good estimate of the risk. For $N$ and $T$ both large, the relevant combination of these parameters turns out to be the aspect ratio $r=N/T$. For $N/T \ll 1$, we will have a good quality estimate. Upon approaching a critical value, which for the optimization of variance is $N/T = 1$, the sample to sample fluctuations become stronger and stronger, until at $N/T = 1$ the average relative estimation error becomes infinitely large \cite{pafka2002Noisy,pafka2003Noisy}. At this point the covariance matrix ceases to be positive definite, and the optimization task becomes meaningless. We can regard $N/T = 1$ as a critical point at which a phase transition is taking place.  

Similar critical phenomena appear also for other risk measures. In the case of ES we have another control parameter, the confidence limit $\alpha$, in addition to the aspect ratio $N/T$.  There is a different critical value of $N/T$ for each $\alpha$ between 0 and 1, thus we have a critical line on the $\alpha$ -- $N/T$ plane. This critical line separates the region where the optimization can be carried out from the region where it is not feasible. We will refer to this line as the phase boundary. In the special case of i.i.d. normal underlying returns, the phase boundary of ES was partially traced out by numerical simulations in \cite{kondor2007Noise} and determined by analytical methods in \cite{ciliberti2007On}. It is displayed in Fig.1.

\begin{figure}[H]
	\centerline{\includegraphics[width=7cm]{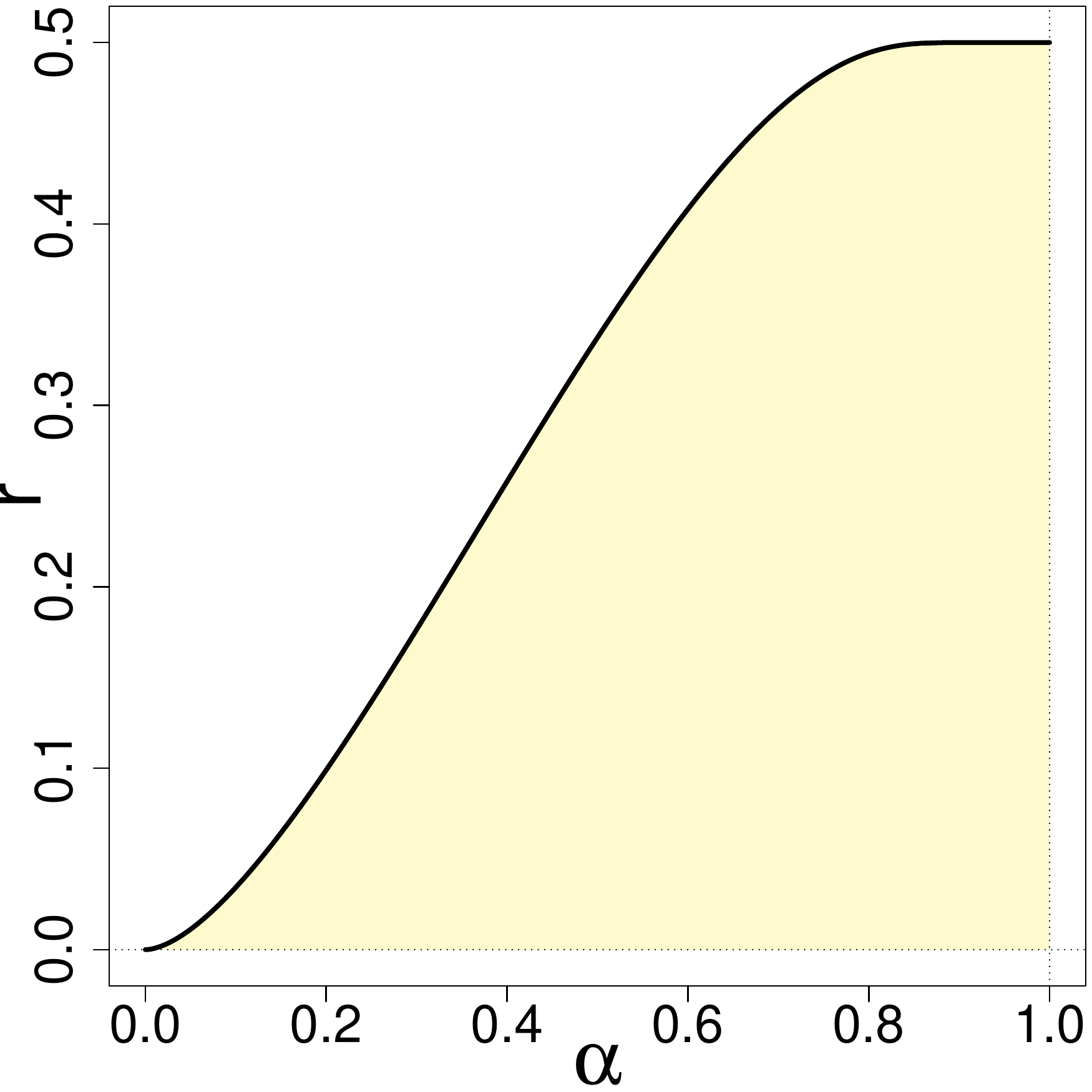}}
	\caption{\footnotesize The phase boundary of ES for i.i.d. normal underlying returns. In the region below the phase boundary the optimization of ES is feasible and the estimation error is finite. Approaching the phase boundary from below, the estimation error diverges, and above the line optimization is no longer feasible.}
	\label{figCriticalLine}
\end{figure}

The phase transition in the optimization of variance has a simple linear algebraic origin: for $T$ smaller than $N$ the rank of the covariance matrix becomes smaller than its dimension. Accordingly, the transition is always sharp: as long as $T$ is larger than $N$, we will always find an optimal portfolio (although in the vicinity of the critical point it may be very far from the ``true'' optimal portfolio), while for $T \le N$ there is no solution. The instability of ES is a little more complicated: it stems from two sources. One is the usual lack of sufficient data, the other is related to the fact that ES as a risk measure is unbounded from below. If in a given sample one of the assets (or a combination of assets) happens to dominate the others (i. e. produces a larger return than any of the others at each time point in the sample) then the investor guided by minimizing ES will be induced to take up a very large long position in the dominating asset and go correspondingly short in the dominated ones, thereby producing an arbitrarily large negative ES, equivalent to an arbitrarily large gain. Of course, the dominance relationship may be entirely due to a random fluctuation and may disappear, or even reverse, in the next sample. This mirage of arbitrage was analysed in detail and shown to be a general property of downside risk measures in \cite{varga2008TheInstability,kondor2010Instability}. It is this mechanism that explains the paradoxical behaviour of the phase boundary, namely that it is sloping down as we go from right to left in Fig.1: the instability of ES occurs at lower and lower values of $N/T$ as we decrease the confidence level, that is as we retain more and more data. Indeed, if we  include not only the far tail, but also the bulk of the distribution, the probability that ES takes a negative value increases, until at $\alpha = 0$ it becomes a certainty, and the phase boundary reaches the horizontal axis.

The earlier studies \cite{ciliberti2007On,varga2008TheInstability,kondor2010Instability,caccioli2013Optimal,caccioli2014Lp} were mainly concerned with the instability of ES, the behaviour of the estimation error in the vicinity of the phase transition, and the possibility of taming the instability via regularization. Here, we will construct the lines along which the estimation error takes up finite, fixed values, i. e. we present a contour map of the estimation error for ES.

A related set of contour lines will also be constructed for a quantity we call the susceptibility, which measures the sensitivity of the estimated ES to a small change in the returns, and yet another quantity that was suggested as a proxy for VaR \cite{rockafellar2000Optimization}.

Some of our results will be presented in a series of figures that can be read as maps of the estimation error landscape. In order to facilitate the quantitative understanding of these results, we will also present numerical results in a tabular form. These tables allow one to determine the minimum amount of data necessary to stay below a stipulated value of estimation error for a given portfolio size and at a given confidence level.

The method behind our analytical results has been described, with minor variations, in \cite{ciliberti2007On,varga2008TheInstability,caccioli2013Optimal,caccioli2014Lp,ciliberti2007Risk}; nevertheless, we will present a brief summary in \ref{sec:appendixA}, for completeness. The salient point is noticing that the task of averaging over statistical samples is analogous to what is called ``quenched averaging'' in the theory of random systems. One can therefore borrow the tools of this theory, in particular the method of replicas \cite{mezard1987Spin}. 

The replica method is very powerful and is capable of delivering results which have not been accessible through any other analytical approach so far. To position our work vis-a-vis the extended literature on the estimation error problem in portfolio selection, we note that most of this literature is based on the analysis of finite (empirical or synthetic) samples, combined with various noise reduction methods ranging from Bayesian \cite{jobson1979Improved,jorion1986Bayes,frost1986AnEmpirical}, through shrinkage \cite{ledoit2003Improved,ledoit2004AWell,ledoit2004Honey,golosnoy2007Multivariate}, lasso \cite{brodie2009Sparse}, random matrix \cite{laloux1999Noise,laloux2000Random,plerou1999Universal,plerou2000ARandom}, and a number of other techniques. The set of purely theoretical papers that deal with the problem with the standard tools of probability theory and statistics is much smaller and concerned with the estimation error in mean-variance optimization \cite{kempf2006Estimating,frahm2010Dominating,okhrin2006Distributional}. A relatively recent review of the field of portfolio optimization is \cite{elton2007Modern}. Analytical results on the estimation error problem of risk measures other than the variance do not exist beyond the few papers \cite{ciliberti2007On,varga2008TheInstability,caccioli2013Optimal,caccioli2014Lp,ciliberti2007Risk} that applied replicas in the present context.   

It should be noted, however, that the method of replicas has a shortcoming in that at a certain point of the derivation one has to analytically continue the formulae from the set of natural numbers to the reals, and the uniqueness of this continuation is very hard to prove. In a number of analogous problems in statistical physics a rigorous proof could be constructed even in the much more complicated case of a non-convex cost function \cite{talagrand2003Spin}. Although we cannot offer such a rigorous proof here, it is hard to imagine how the method could lead one astray when the cost function is convex and has a single minimum. Nevertheless, in the lack of a rigorous mathematical proof we felt compelled to always check the results of replicas by numerical simulations, and always found complete agreement.

The main conclusion this study has led us to is that the error both in the composition of the estimated optimal portfolio and its resulting ES is very large unless the aspect ratio $N/T$ is very small. Qualitatively, this is a foregone conclusion. The novelty is the set of quantitative results showing exactly how large the sample sizes should be to achieve a reasonably low level of estimation error, and how much these sample sizes exceed anything that can be realistically hoped to be available in the industry. It also turns out that including more data (setting a lower confidence level) would not help: the contour lines of relative error in the acceptable range of, say 5\%, are rather flat. What all this means is that for typical parameter values $N, T$ the estimation error is so large as to make portfolio optimization illusory.

Everything we said so far concerns historical estimates. One may expect that parametric estimates suffer less from the estimation error. This is indeed so - though the difference is less than what one might have hoped for. In \cite{varga2008TheInstability} we derived the phase boundary for parametric VaR and ES by the technique of replicas again and found that the critical line of parametric ES lies above that of the historical estimate, so it moves in the good direction. In this paper we extend those results and construct the countour lines of parametric estimates along which ES is a given finite constant. We find that the parametric estimates are indeed less demanding than the historical ones, which is natural, given that with the choice of the target distribution we project a lot of information into the estimate. The gain is, however, far from sufficient to allow even the parametric estimates to produce acceptably low estimation errors for realistic portfolio and sample sizes - and this despite the fact that we were fitting a Gaussian distribution to finite samples generated by a Gaussian. In real life one should fit a fat tailed distribution to empirical data, a task as fraught with uncertainty as estimating a high quantile or a conditional average above it.

Finally, a word on regularization. The standard way of dealing with high-dimensional statistics is to use regularization \cite{buhlmann2011Statistics}, which in the given context would mean imposing a penalty on the large excursions of the portfolio weights,  thereby reducing the estimation error. We studied the effect of regularization on the estimation of ES in \cite{caccioli2013Optimal,caccioli2014Lp,still2010Regularizing}. Here, we refrain from considering the effect of possible regularizers, because our primary purpose is to show up how serious the raw estimation error problem is. We plan to return to the study of various regularizers in a future publication where we wish to assess the bias-estimation error tradeoff in richer data generating processes than the i.i.d. Gaussian considered here.

The plan of the paper is as follows: In Sec. 2 we lay out the task of optimizing ES, fix notation and recall how \cite{rockafellar2000Optimization} reduced this problem to linear programming. In Sec. 3 we define the various quantities characterizing the estimation error: the in-the-sample and out-of-sample estimates of ES, the relative estimation error, the sample average of the estimated distribution of portfolio weights, and the susceptibility. These are the quantities we set out to calculate by the method of replicas. The explanation of the replica method is relegated to \ref{sec:appendixA}, where the generating functional whose minimum will give the answer to our optimization problem is derived as a function of six variables, the so called order parameters. The first order conditions determining the order parameters are written up in Sec. 4, where also the various measures of estimation error defined in the previous section are identified in the replica language. The main features of the solutions of the first order conditions are explained in Sec. 5. The solutions themselves are mainly obtained by numerical computation, in a few special cases one can gain insight into the structure of the equations by analytical calculation, some details of which are presented in \ref{sec:appendixB}. The central results for historical estimates are presented in Sec. 6 mainly in graphical, but also in tabular form. Section 7 discusses the problem of parametric estimates, and makes a comparison with the historical ones. The bulk of the paper is dealing with the simplest possible realization of the estimation error problem: an i.i.d. normal underlying process, estimation of the global minimum portfolio (omitting the constraint of the expected return), etc. In Sec. 8 we consider each of these simplifications in turn, and look into whether they could modify the main message of the paper. Correlated and non identically distributed but still Gaussian underlying fluctuations could be easily accommodated, as could the inclusion of the constraint on expected return. Where numerical simulations remain the only tool are the problems of fat tailed distributions, and the error bars on the average estimation error. These simulations do not pose a problem in principle, but are very computation intensive, so we just present a few illustrative examples. The conclusion of the study of all these possible extensions is that they can modify some of the details of the results obtained in the simplest setup, but do not change the main message of the paper in any meaningful way. Finally, in Sec.9 we summarize the most important results, and indicate the directions along which the present work can be continued.

\section{The optimization of ES}

The simple portfolios we consider here are linear combinations of $N$ securities, with returns ${x_i}, i=1,2,...,N$ and weights $w_i$:

\be
X = \sum_{i=1}^N w_i x_i
\ee

The weights will be normalized such that their sum is $N$, instead of the customary 1. The motivation for choosing this normalization is that we wish to have weights of order unity, rather than $1/N$, in the limit $N\to\infty$:

\be\label{eqBudgetConstraint}
\sum_{i=1}^N w_i = N.
\ee

Apart from this budget constraint the weights will not be subject to any other condition. In particular, they can take any real value, that is we are allowing unlimited short positions. Admittedly, this is rather unrealistic: Depending on the type of institutional investor, short positions may be limited or even excluded by legal and/or liquidity constraints. However, when they are present, even if subject to limits, they greatly contribute to the instability of ES. A ban on short positions would act as a hard $ l_1$ regularizer and would eliminate the instability \cite{caccioli2014Lp}, at least for what concerns the magnitude of ES. (Large fluctuations in the optimal weights may remain even after regularization.) A detailed discussion of the effects of various regularizers will be left for a separate publication, here we focus on the simplest, unregularized  case and wish to display the estimation error stemming from the intrinsic instability of the problem.

We do not impose the usual constraint on the expected return on the portfolio either, so we are looking for the global minimum risk portfolio. This setup is motivated by simplicity. Imposing a constraint on the expected return would not pose any serious difficulty and would not change our conclusions very seriously (only would make them stronger). We will briefly comment on this extension later in the paper.

The probability for the loss $\ell(\{w_i\},\{x_i\})=-X$  to be smaller than a threshold $\ell_0$ is: 

\begin{equation*}
P(\{w_i\},\ell_0) = \int \Pi_{i} d x_i p(\{x_i\}) \theta\left(\ell_0- \ell(\{w_i\},\{x_i\})\right)
\end{equation*}

where $p(\{x_i\})$ is the probability density of the returns, and $\theta(x)$ is the Heaviside function: $ \theta(x) = 1$ for $x> 0$, and zero otherwise. The VaR at confidence level $\alpha $ is then defined as:

\be
{\rm VaR}_\alpha(\{w_i\}) = {\rm min} \{\ell_0 : P(\{w_i\},\ell_0) \ge\alpha\}.
\ee

Expected Shortfall is the average loss beyond the VaR quantile:

\be\label{eqESDefinition}
{\rm ES}(\{w_i\}) = \frac{1}{1-\alpha}\int \Pi_i dx_i p(\{x_i\}) \ell(\{w_i\},\{x_i\}) \theta(\ell(\{w_i\},\{x_i\}) - {\rm VaR}_\alpha(\{w_i\}) ).
\ee

Portfolio optimization seeks to find the optimal weights that make the above ES minimal subject to the budget constraint \eqref{eqBudgetConstraint}. Instead, Rockafellar and Uryasev \cite{rockafellar2000Optimization} proposed to minimize the related function

\be
F_\alpha ( \{w_i\},\epsilon) = \epsilon +\frac{1}{1-\alpha}\int \Pi_i dx_i p(\{x_i\})\left[\ell(\{w_i\},\{x_i\})-\epsilon\right]^+
\ee

over the variable $\epsilon$ and the weights $w_i$:

\be
{\rm ES}(\{w_i\}) = {\rm min}_{\epsilon} F_\alpha ({\{w_i\},\epsilon}),
\ee

where $[x]^+ =(x+|x|)/2$.

The probability distribution of the returns is not known, so one can only sample this distribution, and replace the integral in (4) by time-averaging over the discrete observations. Rockafellar and Uryasev \cite{rockafellar2000Optimization} showed that the optimization of the resulting objective function can be reduced to the following linear programming task: Minimize the cost function

\be\label{eqCostFunction}
E(\epsilon,\{u_t\} )= (1-\alpha) T \epsilon + \sum_{t=1}^T u_t
\ee
under the constraints
\begin{equation*}
u_t  \ge 0~~\forall~t,
\end{equation*}
\be
u_t +  \epsilon+\sum_{i=1}^N x_{it} w_i \ge 0~~\forall~t,
\ee
\begin{equation*}
{\rm and}~~~\sum_i w_i =N.
\end{equation*}

We will have to remember at the end that a multiplicative factor has been absorbed into the definition of the cost function, so the cost function is related to the ES by

\be
{\rm ES} = \frac{E}{(1-\alpha) T} ~~~,
\ee

At this stage we are not yet committed to any particular probability distribution, so the returns can be thought of as drawn from a given model distribution function, or observed in the market. The linear programming task as laid out here will serve as an ``experimental'' laboratory for us: drawing the returns from an arbitrary  distribution we can always determine the optimum by numerical simulations. In the special case when the distribution is Gaussian, we can tackle the problem also by analytical methods.

\section{Estimation error}

Let us first consider the simplest possible portfolio optimization task: assume that the returns $x_{it}$ on asset $i$ at time $t$, $i=1,2,...N$; $t= 1,2,...T$ are i.i.d. standard normal variables, and their number $N$ is fixed, but the number of observations $T$ goes to infinity, so that we are observing the ``true'' data generating process.
The value of the portfolio at time $t$ is:

\be
X_t = \sum_i w_i^{(0)} x_{it},
\ee
where the portfolio weights, denoted as $w_i^{(0)},$ are normalized to $N$ as mentioned before, Eq. \eqref{eqBudgetConstraint}.

If we optimize the convex functional ES over the weights for an infinitely large sample of i.i.d. returns $x_{it}$, the optimal weights will all be equal to unity,  by symmetry:         
\be
w_i^{(0)} = 1 ~~\forall~i. 
\ee

The return on the portfolio averaged over an infinitely long time will be zero:
\be
\langle X_t\rangle = \frac{1}{T}\sum_{it} w_i^{(0)} x_{it}\to 0,~~T\to\infty
\ee
(we denote the time averaging over a given sample by $\langle\ldots\rangle$).

Then the true variance of the portfolio will be

\be
{\sigma_p^{(0)}}^2 = \langle X_t ^2\rangle = \sum_{ij}w_i^{(0)} w_j^{(0)} \frac{1}{T}\sum_t x_{it}x_{jt}=\sum_{ij}w_i^{(0)}\delta_{ij}w_j^{(0)}=\sum_i {w_i ^{(0)}}^2 = N,
\ee

where we made use of the fact that the long time average of the covariance  is:

\be
\lim_{T\to \infty} \frac{1}{T}\sum_t x_{it}x_{jt}=\delta_{ij} = \begin{cases}
                        1 ~~\text{if $i=j$} \\
                        0 ~~\text{if $i\neq j$}
                    \end{cases}
\ee

As a linear combination of Gaussian random variables the portfolio $X_t$ is also a Gaussian random variable.  For independent variables the probability distribution factorizes, so its Expected Shortfall can be easily calculated:

\be\label{eqTrueES}
{\rm ES}_{\alpha}^{(0)} =\frac{{\rm exp}{\left\{-\frac{1}{2}\left(\Phi^{-1}(\alpha)\right)^2\right\}}}{(1-\alpha)\sqrt{2\pi}}~\sigma_p^{(0)} ,
\ee

where $\Phi^{-1}$ is the inverse of the cumulative standard normal distribution     
\begin{equation}
\Phi(x)=\frac{1}{\sqrt{2\pi}}\int_{-\infty}^x e^{-y^2/2} dy
\end{equation}

Eq. (\ref{eqTrueES}) is the true value of ES, the one that would be assigned to an infinitely long stream of the $N$ i.i.d. standard normal returns. 

Let us now pretend that we do not know the true data generating process, and do not have infinitely many observations from which to reconstruct it and deduce the true value of ES.  Instead, we have finite samples of length $T$:

\begin{equation*}
\{x_{it}\},~i=1,2,\ldots,N~;~t=1,2,\ldots,T,
\end{equation*}

and the finite sample average return 

\be
\langle X_t\rangle = \frac{1}{T} \sum_{it} w_i x_{it}
\ee

will depend on the sample. If we optimize ES as a functional of the variables $x_{it}$ over a finite sample, the optimal weights will not all be equal, they will display a certain distribution around their true value of 1. This distribution will be different in the different samples. Let us denote the average over the samples by an overbar. Then the sample average of the return $\langle X_t\rangle$ will be
\be
\overline{\langle X_t\rangle} = \sum_i\frac{1}{T}\overline{\sum_t w_i x_{it}}\equiv\sum_i\overline{\langle w_i x_{it}\rangle}.
\ee

Here both the optimal weights and the returns depend on the sample, as such they are not independent of each other. However, by symmetry, the average $\frac{1}{T}\overline{\sum_t w_i x_{it}}$ will be independent of $i$, so

\be
\overline{\langle X_t\rangle} = N\overline{\langle wx\rangle}.
\ee

The variance of the portfolio return in a given sample

\be\label{eqSampleVariance}
\langle X_t^2\rangle - \langle X_t\rangle ^2 =\sum_{ij} w_i w_j \left( \frac{1}{T}\sum_t x_{it} x_{jt}-\frac{1}{T}\sum_t x_{it} \frac{1}{T}\sum_t x_{jt}\right)
\ee

is also a random variable. Its sample average is

\be\label{eqSampleAverage}
\sigma_{p,in}^2 = \overline{\langle X_t^2\rangle-\langle X_t\rangle^2}\equiv\overline{\sum_{ij} w_i w_j C_{ij}},
\ee

where $C_{ij}$ is the covariance matrix of the returns in a given sample. The weights in (\ref{eqSampleVariance}) are supposed to be those that are optimized under the convex risk measure ES within a given sample, and $C_{ij}$ is the estimated covariance matrix in that sample. Therefore  \eqref{eqSampleVariance} gives the in-the-sample estimate of the portfolio variance, and the in-the-sample standard deviation $\sigma_{p,in}$ multiplied by ${\rm exp}{\left\{-\frac{1}{2}\left(\Phi^{-1}(\alpha)\right)^2\right\}}/(1-\alpha)\sqrt{2\pi}$, gives the in-the-sample estimate of ES, while  \eqref{eqSampleAverage} gives the sample average of the in-the-sample estimate of the portfolio variance. In-the-sample estimates can, however, be grossly misleading, especially near a critical point where sample to sample fluctuations are large. The relevant measure of estimation error is the out-of-sample estimate of the variance, where the weights are still those optimized within the sample, but the covariance matrix is the true covariance matrix $\delta_{ij}$ of the process. Thus the out-of-sample variance of the portfolio will be:

\be\label{eqOutOfSampleVariance}
\sigma_{p,out}^2 = \overline{\sum_{ij}w_i w_j\delta_{ij}}=\overline{\sum_{i}w_i^2}= N\overline{w^2}.
\ee

This quantity is directly related to the variance of the weights distribution. As mentioned above, the estimated values of the weights in finite size samples are different from their true value 1. The variance of the weights distribution averaged over the samples will be:

\be\label{eqWeightVariance}
\sigma_w^2 = \frac{1}{N} \sum_i \left( \overline{w_i^2}-(\overline{w_i})^2\right) = \frac{1}{N}\sum_i \overline{\langle w_i^2\rangle} -1
\ee

where use has been made of the fact that, although the individual weights in a given sample can strongly deviate from their true value of 1, their sample average is still 1.

From \eqref{eqOutOfSampleVariance} and \eqref{eqWeightVariance} we get the relationship between the out-of-sample variance of the portfolio and the variance of the weights:

\be
\sigma_{p,out}^2 = N(\sigma_w^2+1).
\ee

The corresponding formula for the out-of-sample estimate of ES averaged over the samples is:

\be\label{eqOutOfSampleES}
{\rm ES}_{out} = \frac{{\rm exp}{\left\{-\frac{1}{2}\left(\Phi^{-1}(\alpha)\right)^2\right\}}}{(1-\alpha)\sqrt{2\pi}} (N\overline{w^2})^{1/2}.
\ee

A natural measure of the estimation error is the ratio of the estimated ES (\ref{eqOutOfSampleES}) and its true value given in (\ref{eqTrueES}):
\be
\label{eqTrueES0}
\frac{{\rm ES}_{out}}{{\rm ES}^{(0)}} = (\overline{w^2})^{1/2}=(\sigma_w^2+1)^{1/2}.
\ee

This ratio is always larger than one. Subtracting 1 we obtain the relative estimation error of ES:

\be
\frac{{\rm ES}_{out}}{{\rm ES}^{(0)}} -1= (\sigma_w^2+1)^{1/2}-1.
\ee

If the sample size $T$ is very large relative to the number of different assets $N$, that is when the aspect ratio $r=N/T$ is small, we do not expect large fluctuations in the weights, so $\sigma_w^2$ as well as the estimation error $\frac{{\rm ES}_{out}}{{\rm ES}^{(0)}}$ will be small. In the opposite case, when the sample size is not sufficiently large (and from the phase diagram in Fig. 1 we know that for small confidence levels this may happen already for small $r$'s), there will be violent fluctuations in the weights with very large short positions compensated by very large long ones. As a result, the variance of the weight distribution as well as the relative estimation error in ${\rm ES}$ will be very large, ultimately diverging at the phase boundary.

The importance of the distribution of weights in characterizing the estimation error was suggested to one of us (I.K.) by Sz. Pafka (private communication) several years ago. 

It is an interesting question how sensitive the estimation error is to small variations in the returns. We will consider the simplest such variation: a uniform shift of all the returns by a small amount: $x_{it} ~\to ~ x_{it} + \xi $. This will cause a change in the estimated optimal weights. We wish to characterize the sensitivity of the estimation error by the derivative with respect to $\xi$ of the expression in (\ref{eqTrueES0}), taken at $\xi = 0$. We call this quantity the susceptibility and denote it by $\chi$:

\be
 \label{susceptibility}
  \chi = \frac{\partial}{\partial\xi} \left( \frac{{\rm ES}_{out}}{{\rm ES}^{(0)}} \right)_{\xi=0} =  \frac{\partial}{\partial\xi}  \left( \overline{w_i^2} \right)^{1/2}_{\xi=0}
\ee

The analytical treatment to be presented in the next section will provide the distribution of weights, the in-the-sample and the out-of-sample estimates for Expected Shortfall, as well as the susceptibility in the limit of large $N$ and $T$, with their ratio $r = N/T$ fixed. As a bonus, we will also obtain results for a quantity that was suggested in \cite{rockafellar2000Optimization} to be a proxy for the estimated VaR and which is, in fact, the VaR of a portfolio optimized under ES.

\section{The first order conditions}
 As mentioned earlier, the analytical solution to the optimization problem (\ref{eqCostFunction}) can be found for Gaussian returns in the limit of large $N$ and $T$ by methods taken over from the statistical physics of random systems. The method has been explained in                                 \cite{ciliberti2007On,varga2008TheInstability,caccioli2013Optimal,caccioli2014Lp,ciliberti2007Risk}, but we include the main points of the derivation in \ref{sec:appendixA}, for completeness.  The essence of the method is the following: the cost function is regarded as the Hamiltonian (energy functional) of a fictitious statistical physics system, a fictitious temperature is introduced and the free energy (the logarithmic generating function) of this system is calculated in the limit $N,T\to\infty$ with $N/T= r$ fixed. The original optimization problem is recovered in the limit when the fictitious temperature goes to zero. Averaging over the different random samples of returns corresponds to what is called quenched averaging in statistical physics. We take up the discussion from Eq.(\ref{freeEnergy})         
where the cost function has already been averaged over the samples and is expressed as the function of a much reduced number of variables (from the $N+T+1$ in (\ref{eqCostFunction}), 
down to six), the so-called order parameters, as follows:
\bea
\label{free_energy}
F({\lambda},{\epsilon},{q}_0,\Delta, {\hat{q}}_0,\hat{\Delta})&=&
{\lambda} +\tau (1-\alpha)\epsilon -\Delta{\hat{q}}_0-\hat{\Delta}{q}_0\\
\nonumber &+& \langle {\rm min}_w \left[V(w,z)\right]\rangle_z
 +\frac{\tau\Delta}{2\sqrt{\pi}}\int_{-\infty}^{\infty}ds e^{-s^2}
g\left({\frac{\epsilon}{\Delta}}+s \sqrt{\frac{2{q}_0}{\Delta^2}} \right),
\eea
where  
\begin{equation}
\label{pot}
V(w,z)=\hat{\Delta} w^2 -{\lambda} w -z w\sqrt{-2{\hat{q}}_0}.
\end{equation}

and
\begin{equation}
    g(x)= \left\{ \begin{array}{cc} 0 ,&    x\ge 0\\
    x^2 , &  -1\le x\le 0\\
    -2 x-1, & x<-1
     \end{array} \right..
\end{equation}

In (\ref{free_energy})  $\langle\cdot\rangle_z$ represents an average over the standard normal variable $z$.

The value of the free energy, i.e. the minimal cost per asset,  is ultimately a function of the two control parameters, the aspect ratio $r=N/T$ and the confidence limit $\alpha$. In order to find this function, one has to determine the minimum of the above expression in the space of the six order parameters ${\lambda},{\epsilon},{q}_0,\Delta, {\hat{q}}_0$ and $\hat{\Delta}$, find the values of these as functions of the control parameters, and substitute them back into (\ref{free_energy}). 

We will see below that of the six order parameters three can easily be eliminated, so we end up with three equations for the three remaining order parameters, in accord with the setup in \cite{ciliberti2007On} where the replica method was first applied in a portfolio optimization context. Thus it may seem that our present approach, with its six order parameters and the nested optimization structure in (\ref{free_energy}) is making  an unnecessary detour. This is not quite so: the present scheme allows us to deduce along the way, in addition to the optimal cost, also the sample averaged distribution of the estimated optimal portfolio weights and the susceptibility, i.e. a measure of the sensitivity of the weights to changes in the distribution of returns.

Let us now start with the solution of the inner optimization problem in (\ref{free_energy}). It arises from the optimization over the weights $w_i$ in the original problem and the Gaussian random variable $z$ encodes the effect of the randomness in the sample. The solution of this problem gives $w^*(z)$ that we called the ``representative'' weight in \cite{caccioli2014Lp}:

\begin{equation}\label{wstar}
    w^*(z)=  \frac{z\sqrt{-2\hat{q}_0}+\lambda}{2\hat{\Delta}}   
\end{equation}

The sample average of $w^*(z)$ is then
\be
 \label{wstaraverage}
  \langle w^* \rangle_z = \frac{\lambda}{2 \hat{\Delta}}
\ee

while the average of its square is
\be
 \label{wstarsquareaverage}
    \langle {w^*}^2 \rangle_z = \frac{\lambda^2 - 2 \hat{q_0}}{4 \hat{\Delta}^2}
\ee

The probability density of the portfolio weights $p(w)=\langle \delta(w-w^*(z))\rangle_{z}$ ($\delta$ is the Dirac distribution) works out to be a Gaussian centered on $ \langle w^* \rangle_z$
with variance
\be
 \label{weightsvariance}
   \sigma_w^2 =   \langle {w^*}^2 \rangle_z -  \langle w^* \rangle_z ^2 = - \frac{\hat{q_0}}{2 \hat{\Delta}^2}
\ee

Now we spell out the first order conditions that determine the order parameters:
 
\be
1=\avg{w^*}_{z}\label{spBudget}
\ee
\be
\label{EqFirstOrder2}
(1-\alpha)+\frac{1}{2\sqrt{\pi}}\int_{-\infty}^\infty ds e^{-s^2}g'\left(\frac{\epsilon}{\Delta}+s\sqrt{\frac{2{q}_0}{\Delta^2}}\right)=0
\ee
\be
\hat{\Delta} - \frac{1}{2r\sqrt{2\pi{q}_0}}\int_{-\infty}^\infty ds e^{-s^2}s g'\left(\frac{\epsilon}{\Delta}+s\sqrt{\frac{2{q}_0}{\Delta^2}}\right)=0
\ee

\be
-{\hat{q}}_0-2\frac{\hat{\Delta}{q}_0}{\Delta} + \frac{1}{2r\sqrt{\pi}}\int_{-\infty}^\infty ds e^{-s^2}g\left(\frac{\epsilon}{\Delta}+s\sqrt{\frac{2{q}_0}{\Delta^2}}\right)+\frac{(1-\alpha)}{r}\frac{{\epsilon}}{\Delta}=0
\ee

\be
\Delta=\frac{1}{\sqrt{-2{\hat{q}}_0}}\avg{w^* z}_{z}\label{spDelta}
\ee

\be
{q}_0=\avg{{w^*}^2}_{z}.\label{spQ}
\ee

The first of these stems from the budget constraint and says that the expectation value of the estimated optimal weights averaged over the random samples is just 1. This is an obvious result: the distribution of weights in the random samples can be very different from their true distribution (all equal to 1), but on average they still fluctuate about their true value. On the other hand, from the inner optimization problem we  
found (\ref{wstaraverage}), so we have

 \be
   \label{lambdaDeltahat} 
    \lambda = 2 \hat{\Delta}.
 \ee

 Multiplying (\ref{wstar}) by $z$ and averaging over the random variable $z$ we find that $\langle w^* z\rangle = \sqrt{-2\hat{q_0}} / 2\hat{\Delta}$. Plugging this expression into equation \eqref{spDelta} we get 
 
 \be
  \label{DeltaDeltahat} 
   \Delta = \frac1{2\hat{\Delta}}.
 \ee

 Finally, the average squared weight (\ref{wstarsquareaverage}) is, by the last of the first order conditions, equal to $q_0$, so that we have
 
 \be
  \label{qo through qohatDeltahat}
   q_0 = \frac{\lambda^2 - 2\hat{q_0}}{4 \hat{\Delta}^2} = 1 - \frac{\hat{q_0}}{2 \hat{\Delta}^2},
 \ee

which immediately links $q_0$ to the variance of the weights distribution:

\be
 \label{qosigmaweights}
   q_0 = 1 + \sigma_w^2 \, .
\ee

With this we have used three of the first order conditions to express $\lambda$, $\hat{\Delta}$ and $\hat{q_0}$ through the order parameters $\epsilon$, $\Delta$ and $q_0$, which allows us to eliminate the former group of variables in favour of the latter. We will see shortly that the retained variables $\epsilon$, $\Delta$ and $q_0$ all have a direct meaning. We have also found a useful relationship between the order parameters and the variance of the weights distribution which tells us that the phase boundary should be defined as the line along which $q_0$ or, equivalently, $\Delta$ diverges (or $\hat{\Delta}$ vanishes), because this is the line along which the variance of the weigths goes to infinity, corresponding to a situation where the weights fluctuate wildly, taking up large positive as well as negative values.

The cost function itself can now be found by evaluating the potential $V$ at the optimum
\begin{equation*}
  \avg{V^*}_z = - \hat{\Delta} \avg{{w^*}^2} = - \hat{\Delta} q_0 \,.
\end{equation*}

This, together with (\ref{pot}), (\ref{lambdaDeltahat}) - (\ref{qo through qohatDeltahat})  yields the remarkably simple result $F = 1/\Delta$. Remembering that $F$ is the cost per asset, so  the cost itself is $E=NF$ and also recalling that the cost function has to be divided by $(1-\alpha)T$ in order to get the Expected Shortfall we have:

\be
\label{equationESCost}
  {\rm ES} = F r / (1-\alpha) =\frac{r}{(1-\alpha) \Delta} \,.
\ee

Since we have been optimizing over all the variables to get this expression, this is the in-the-sample estimate of the Expected Shortfall. 

In order to find the out-of-sample estimate, we have to recall (\ref{eqTrueES0}), where the out-of-sample estimate of ES is expressed through the variance of the estimated portfolio weights. Using the result (\ref{qosigmaweights}) for the latter we find:
\be
  \frac{{\rm ES}_{out}}{{\rm ES}^{(0)}} = \sqrt{\langle w^2 \rangle} = \sqrt{1 + \sigma_w^2} = \sqrt{q_0}
\ee

This relationship gives us the meaning of the variable $q_0$: $\sqrt{q_0} -1$ is the relative estimation error of the out-of-sample estimate for ES.

In order to find the meaning of $\Delta$, we consider a small shift in the returns, as in the previous section: $x_{it} ~\to ~ x_{it} + \xi $.  It is easy to see that for such a modified setup the whole derivation of the cost function goes through as before, with the only change that wherever we had $\lambda$ we will have $\lambda$ shifted by $\xi$ as  $\lambda \to \lambda + \xi$. Accordingly, the sample average of the optimal weight will become:
\begin{equation*}
  \langle w^* \rangle_z = \frac{\lambda+\xi}{2 \hat{\Delta}} = (\lambda + \xi) \Delta
\end{equation*}
 
 and its response to the small perturbation $\xi$:
 \be
 \label{equationDeltaMeaning}
   \left. \frac{\partial  \langle w^* \rangle_z}{\partial\xi} \right|_{\xi=0} = \Delta
 \ee

 The same for the average weights squared is
\begin{equation*}
  \langle {w^*}^2 \rangle_z = \frac{(\lambda+\xi)^2 - 2 \hat{q_0}}{4 \hat{\Delta}^2} 
\end{equation*}

 with its response:
  \be
   \left. \frac{\partial  \langle {w^*}^2 \rangle_z}{\partial\xi} \right|_{\xi=0} = \frac{\lambda}{2 \hat{\Delta}^2} = \frac1{\hat{\Delta}} = 2 \Delta \, .
 \ee

  Finally the susceptibility introduced in (\ref{susceptibility}) works out to be 
 \be
  \label{susceptibilityofES}
   \chi = \frac{\partial}{\partial\xi} \left(  \frac{{\rm ES}_{out}}{{\rm ES}^{(0)}} \right)_{\xi=0} =  \frac{\partial}{\partial\xi} \sqrt{q_0} = \frac{\Delta}{\sqrt{q_0}} \, .
 \ee

 Thus, $\Delta$ measures the sensitivity of the weights to small shifts in the returns, and the ratio $\Delta/\sqrt{q_0}$ that appears all throught the first order conditions is the sensitivity of the relative error of the estimated ES.
 
 The third  order parameter is $\epsilon$. This variable was suggested to be a proxy for VaR by Rockafellar and Uryasev \cite{rockafellar2000Optimization}. Indeed, from the setup of the linear programming task, it is obvious that   $\epsilon$ is indeed equal to VaR - the VaR of the portfolio optimized under ES. (We have checked this identification by numerical simulations at several aspect ratios $r$ and confidence levels $\alpha$.) 
 
We make a little digression here, to establish contact with earlier work. As already mentioned, the first paper applying replica methods in a portfolio optimization context \cite{ciliberti2007On} used a three-parameter optimization. The correspondence between that paper and the present one is the following: the order parameter called $q_0$ in \cite{ciliberti2007On} is $q_0/\Delta^2$ here; the variable $v$ there is $\epsilon/\Delta$ here; and the variable $t$ there is the reciprocal of our $r$. With this replacement the cost function there becomes identically equal to the one here. The use of the scaled variables in \cite{ciliberti2007On} was well-justified by the fact that near the phase boundary all three order parameters diverge, and it is the scaled variables that remain finite. Our present interest is wider: we want to solve the problem on the whole $\alpha-r$ plane, and the scaling that is expedient near the phase boundary may not be very useful elsewhere. For example, if $r$ goes to zero (i.e. the sample sizes $T$ is much larger than the dimension $N$) the distribution of weights will be sharp, so $q_0$ will go to 1, while $\Delta$ will vanish. As for $\epsilon$ it will take up the simple form $\Phi^{-1}(\alpha)$ corresponding to the VaR of a sample of Gaussian returns.

Having learned the financial meaning of our three order parameters $q_0$, $\Delta$ and $\epsilon$, we have to turn now to the solution of the remaining three first order conditions, to obtain the order parameters as functions of $r$ and $\alpha$. The first task is to eliminate the variables with a hat through the relationships above. Next we want to get rid of the integrals in the equations. This can be achieved by repeated integration by part. The resulting set of equations is much more amenable to numerical and, in some exceptional cases, analytical solutions. They are as follows:
\be
 \label{equationPhi}
  r = \Phi\left( \frac{\Delta + \epsilon}{\sqrt{q_0}} \right) - \Phi\left( \frac{\epsilon}{\sqrt{q_0}} \right)
\ee

\be
 \label{equationPsi}
  \alpha =  \frac{\sqrt{q_0}}{\Delta} \left\{ \Psi\left( \frac{\Delta + \epsilon}{\sqrt{q_0}} \right) - \Psi\left( \frac{\epsilon}{\sqrt{q_0}} \right)  \right\}
\ee

\be
 \label{equationW}
 \frac1{2\Delta^2} + \frac{\alpha}r \frac{\epsilon}{\Delta} + \frac12 \frac{q_0}{\Delta^2} + \frac1{2 r} = \frac1r \frac{q_0}{\Delta^2}  \left\{ W\left( \frac{\Delta + \epsilon}{\sqrt{q_0}} \right) - W\left( \frac{\epsilon}{\sqrt{q_0}} \right) \right\} \, .
\ee

where
\be
 \label{definitionPhi}
   \Phi(x) = \frac1{\sqrt{2\pi}} \int_{-\infty}^{x} \! dt\, e^{-t^2/2}
 \ee

\be
  \label{definitionPsi}
   \Psi(x) = x \Phi(x) + \frac1{\sqrt{2\pi}}  e^{-x^2/2}
\ee

\be
  \label{definitionW}
   W(x) = \frac{x^2+1}2 \Phi(x) + \frac{x}2  \frac1{\sqrt{2\pi}}  e^{-x^2/2} \, .
\ee

These functions are closely related to each other:
\be
  \Psi^\prime(x) = \Phi(x) \,, \qquad W^\prime(x) = \Psi(x) \, .
\ee

They also exhibit simple symmetries upon changing the sign of the argument:
\bea
 \Phi(x) &=& 1- \Phi(-x) \nonumber \\
 \label{equationSymmetries}
  \Psi(x) &=& x + \Psi(-x) \\
  W(x) &=& \frac{x^2+1}2 - W(-x) \, . \nonumber
\eea

Note that the two equations (\ref{equationPhi}) and (\ref{equationPsi}) are closed for the two ratios formed by the unknowns, and can therefore be solved for them independently of the third equation. The third equation, (\ref{equationW}) will determine $\Delta$ separately, and with that the other two unknowns as well.

\section{Solution of the first order conditions}

The set of equations~(\ref{equationPhi}) -(\ref{equationW}) is nontrivial, the solutions become singular along the phase boundary, moreover, at the two endpoints  $r=\alpha=0$ and $r=1/2, \alpha=1$ the solutions have essential singularities, with the limits depending on the direction from where we approach these points, while the solutions are non-analytic all along the $\alpha=1$ line. Nevertheless, it is possible to gain a first orientation along some special lines by analytical calculations.

The most obvious case is the interval $0<\alpha<1$ on the horizontal axis. This corresponds to $r=N/T\to 0$ , that is to a situation where we have much more observations than the dimension of the portfolio. Here the distribution of the weights must be sharp (all weights equal to 1), so the variance of this distribution must be zero, which implies $\Delta=0$ and $q_0=1$. At the same time, $\epsilon$ must be the VaR of an i.i.d. normal portfolio, i.e. $\Phi^{-1}(\alpha)$. It is easy to see that this triplet is indeed a solution of the first order conditions along the horizontal axis. Note that $\epsilon$ is zero at $\alpha=1/2$ and positive resp. negative on the right resp. left of this point. Furthermore $\epsilon$ diverges at both ends, going to $-\infty$ and $+\infty$ at $\alpha=0$ and $\alpha=1$, respectively. One can make an expansion assuming $r$ small; this works for all $\alpha$, except the two end points. 

Another analytically tractable case is that of the vertical interval $\alpha = 1,~ 0<r<1/2$. One can show that $\epsilon$ and $q_0$ are finite, while $\Delta$ diverges here. The interest in this case stems from the fact that $\alpha=1$ corresponds to the minimax risk measure introduced in \cite{young1998AMinimax}, the best combination of the worst losses. Again, an expansion can be made in the vicinity of this vertical line, except at the two endpoints $r=0, \alpha=1$ and $r=1/2, \alpha=1$.

Two further special lines along which one can make analytical progress are the vertical line at $\alpha=1/2$ and the one along which $\epsilon=0$, the latter running from $\alpha=1/2, r=0$ to $\alpha=1, r=1/2$.

The most important special line is the phase boundary shown in Fig. 1. All three order parameters diverge at this line, but their ratios stay finite. This line was analytically derived in \cite{ciliberti2007On} where also the nature of the divergence and the scaling near the critical line were explored. The root of this divergence has been identified in \cite{varga2008TheInstability,kondor2010Instability} as the apparent arbitrage arising from the statistical fluctuations in finite samples.

The intricacies of the first order conditions are not our primary focus here, so further details are relegated to ~\ref{sec:appendixB}. The rest of this Section is devoted to the presentation of the numerical solutions of the first order conditions. The results will be displayed in the form of a few contour maps, i.e. sets of lines along which the order parameters are constant. These contour plots should be read as the maps of a landscape, the parameters on the lines are the fixed values of the functions that are being plotted in the figure.

\begin{figure}[H]
  \centering
  \begin{minipage}[t][][t]{65mm}
    \includegraphics[width=65mm]{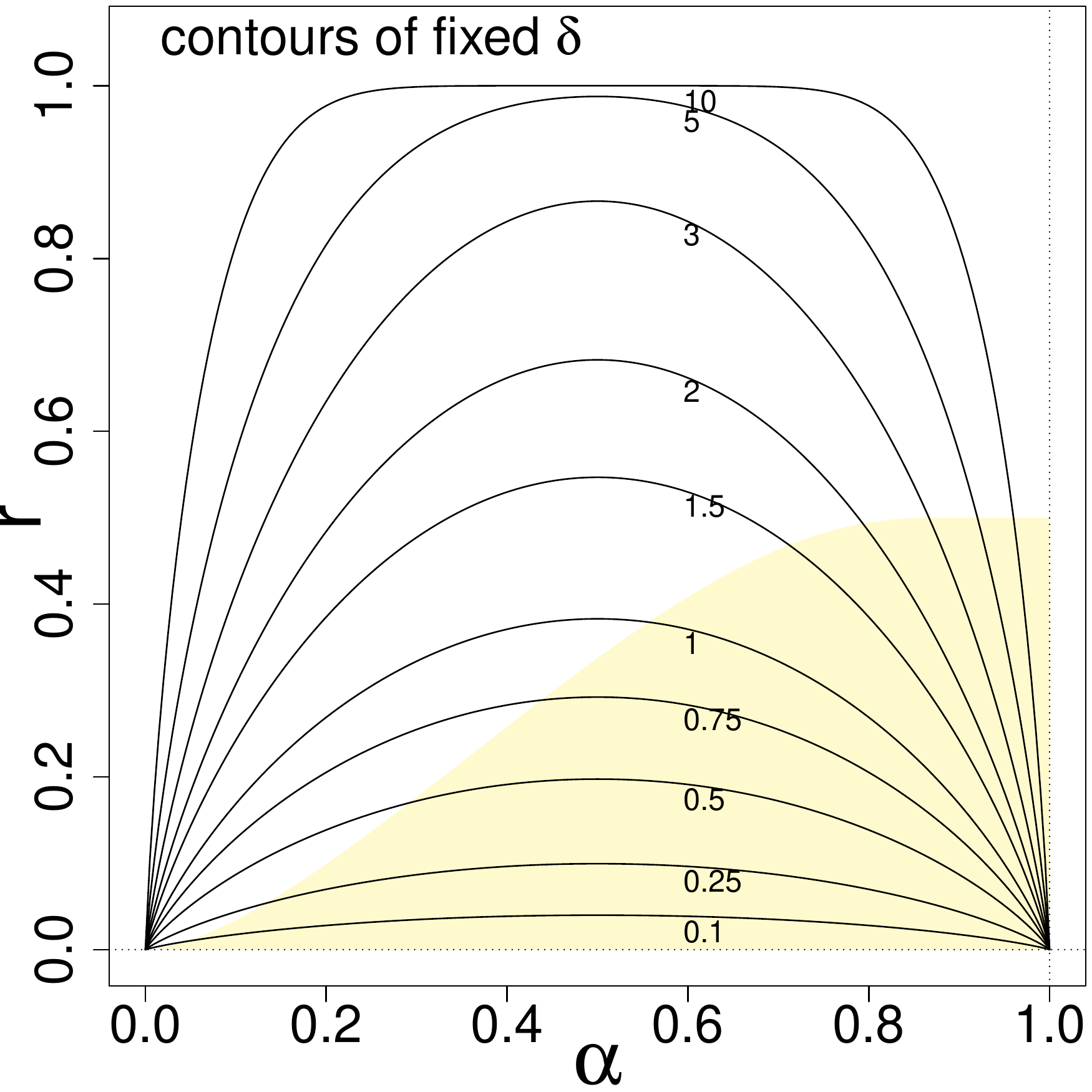}
    \caption{\footnotesize Contour map of the ratio $\frac{\Delta }{\sqrt{q_0}} = \delta$ measuring the sensitivity of the relative estimation error of ES to small changes in the returns. The value of $\delta$ is the parameter on the curves; $\delta$ is constant along these lines. As $\delta$ is increasing, the curves fill in the unit square.}
    \label{fig:smalldelta}
    \end{minipage}
  \qquad
  \begin{minipage}[t][][t]{65mm}
    \includegraphics[width=65mm]{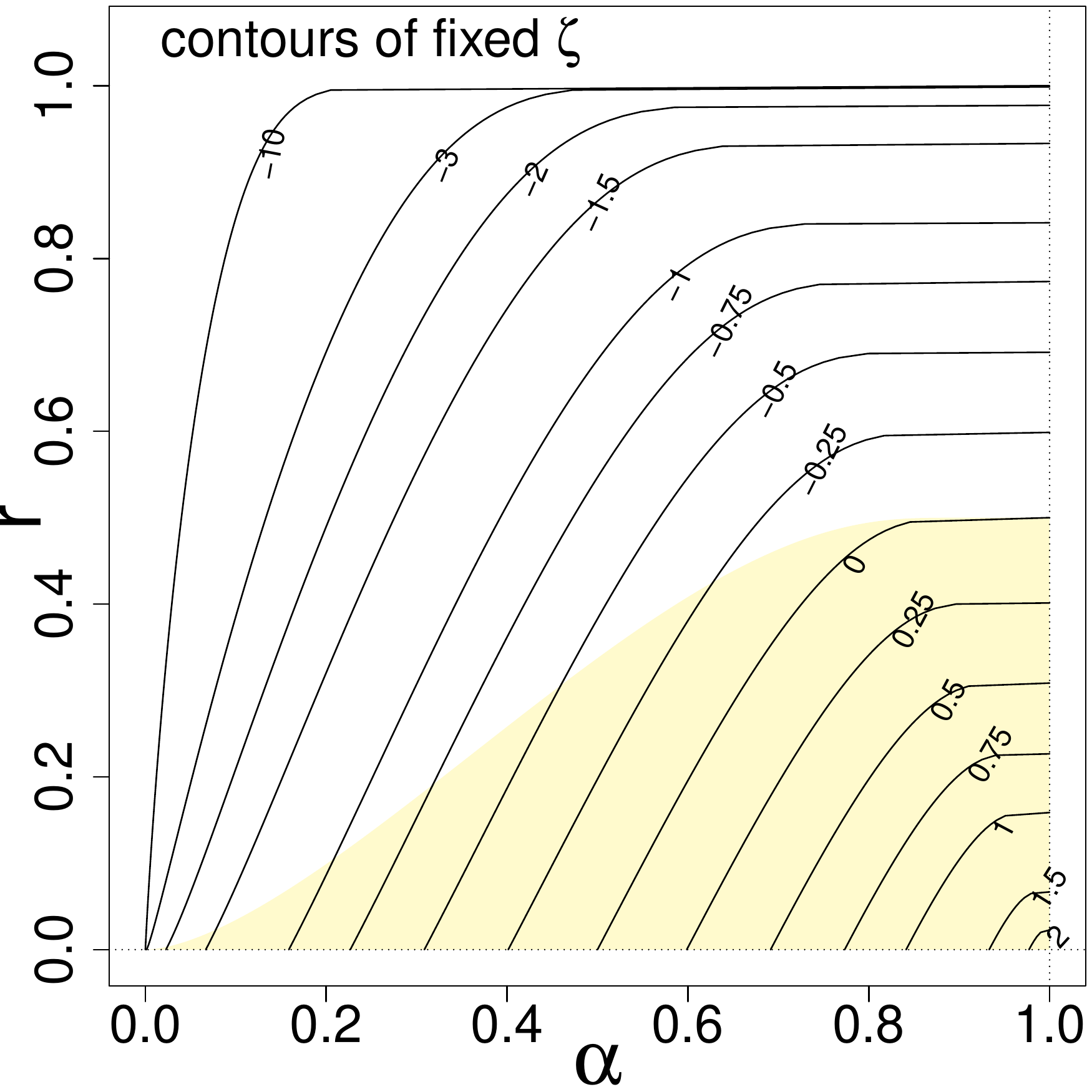}
    \caption{\footnotesize Contour map of the ratio $\frac{\epsilon}{\sqrt{q_0}} =\zeta$. As $\zeta$ sweeps the range ($-\infty$ , $+\infty$) the lines of constant  $\zeta$ fill in the unit square.}
    \label{fig:zeta}
  \end{minipage}
\end{figure}

As we have already noted, the structure of the set of equations~(\ref{equationPhi}-\ref{equationW}) is such that the first two determine the ratios $\frac{\Delta }{\sqrt{q_0}}$ and $\frac{\epsilon}{\sqrt{q_0}}$. The solutions for these ratios are presented in Figs.~\ref{fig:smalldelta} and \ref{fig:zeta}. These ratios remain finite as we cross the phase boundary, so they can be continued beyond the feasible region (shown as the shaded area in Figs.~\ref{fig:smalldelta},\ref{fig:zeta}). The $\frac{\Delta}{\sqrt{q_0}}$ contour lines display a symmetry that goes back to the symmetries of the functions $\Phi$ and $\Psi$ given in~(\ref{equationSymmetries}). These lines all tend to the point $r=0, \alpha=1$, falling off steeper and steeper as we go to higher values of the parameter on the lines. Assume we erect a vertical line at some $\alpha_0$ close to 1, say, at the confidence level $\alpha=0.975$ favoured by regulation. If we now choose a very small $r$, the point $r, \alpha_0$ will fall on a curve corresponding to a relatively small value of $\frac{\Delta}{\sqrt{q_0}}$. This ratio has been identified as the susceptibility, the sensitivity of the estimated $\textrm{ES}_{out}$ to small changes in the returns. A small value of the susceptibility means our estimate is rather stable against changes in the observed prices. As we move upwards along the $\alpha=\alpha_0$ line, that is as we are considering larger and larger $r$'s (shorter and shorter time series), the susceptibility grows very fast: if we do not have enough data, our estimate will be extremely sensitive to price changes.

It is remarkable that right at $\alpha=1$, the susceptibility is infinitely large for any finite $r$: the minimax problem is infinitely sensitive to any change in the returns. This is plausible: if we are taking into account only the worst outcomes, our estimated risk measure will be shifted even by an infinitesimal price change.

Let us turn now to Fig.~\ref{fig:zeta}. It shows the contour lines of $\frac{\epsilon}{\sqrt{q_0}}$. As can be seen, the curves corresponding to positive $\epsilon$'s all bend over and hit the $\alpha=1$ line inside the feasible region, whereas the negative $\epsilon$ curves cross the phase boundary and reach the $\alpha=1$ line between the critical point at $r=1/2$ and $r=1$ that lies in the unfeasible region.

There is a remarkable feature showing up in both Figs.~\ref{fig:smalldelta} and \ref{fig:zeta}. If we allow the ratio $\frac{\Delta}{\sqrt{q_0}}$ to go from zero all the way up to infinity, the resulting contour lines will fill the whole unit square. Likewise, as $\frac{\epsilon}{\sqrt{q_0}}$ goes from minus infinity to plus infinity, the contour lines will fill the unit square again, but neither of these sets of curves ever go beyond $r=1$. We have to remember that we are considering a situation here such that both $N$ and $T$ are infinitely large with a fixed ratio $r$, and the phase boundary was derived in this particular limit. In the special case of the minimax problem ($\alpha=1$), the feasibility or otherwise of optimizing ES can, however, be decided also for finite $N$ and $T$. For finite $N$ and $T$ there is no sharp phase boundary (there is no phase transition in a finite system), instead the probability that the optimization can be carried out is high, but less than 1 for $N/T<1/2$, small, but non-zero for   $1/2 < N/T < 1$, and identically zero for $N/T>1$ \cite{kondor2007Noise}. If  $N$ and $T$ go to infinity with their ratio $r=N/T$ kept finite, the high probability for $r<1/2$ becomes 1, the small probability for $1/2<r$ becomes zero, so the critical point gets pinned at $r=1/2$. The behaviour of the $\frac{\Delta}{\sqrt{q_0}}$  and $\frac{\epsilon}{\sqrt{q_0}}$  curves suggests that for finite $N$ and $T$ a similar scenario is to be expected for any $\alpha$ between zero and one: we conjecture that if one were able to generalize the combinatorial result in \cite{kondor2007Noise} from $\alpha=1$ to a generic confidence level, one would find a solution with high probability in the region which ultimately becomes the feasible region for $N,T\to\infty$ , with small probability above the to-be phase boundary, and zero probability above $r=1$. 
\begin{figure}[h]
  \centering
  \begin{minipage}[t][][t]{65mm}
    \includegraphics[width=65mm]{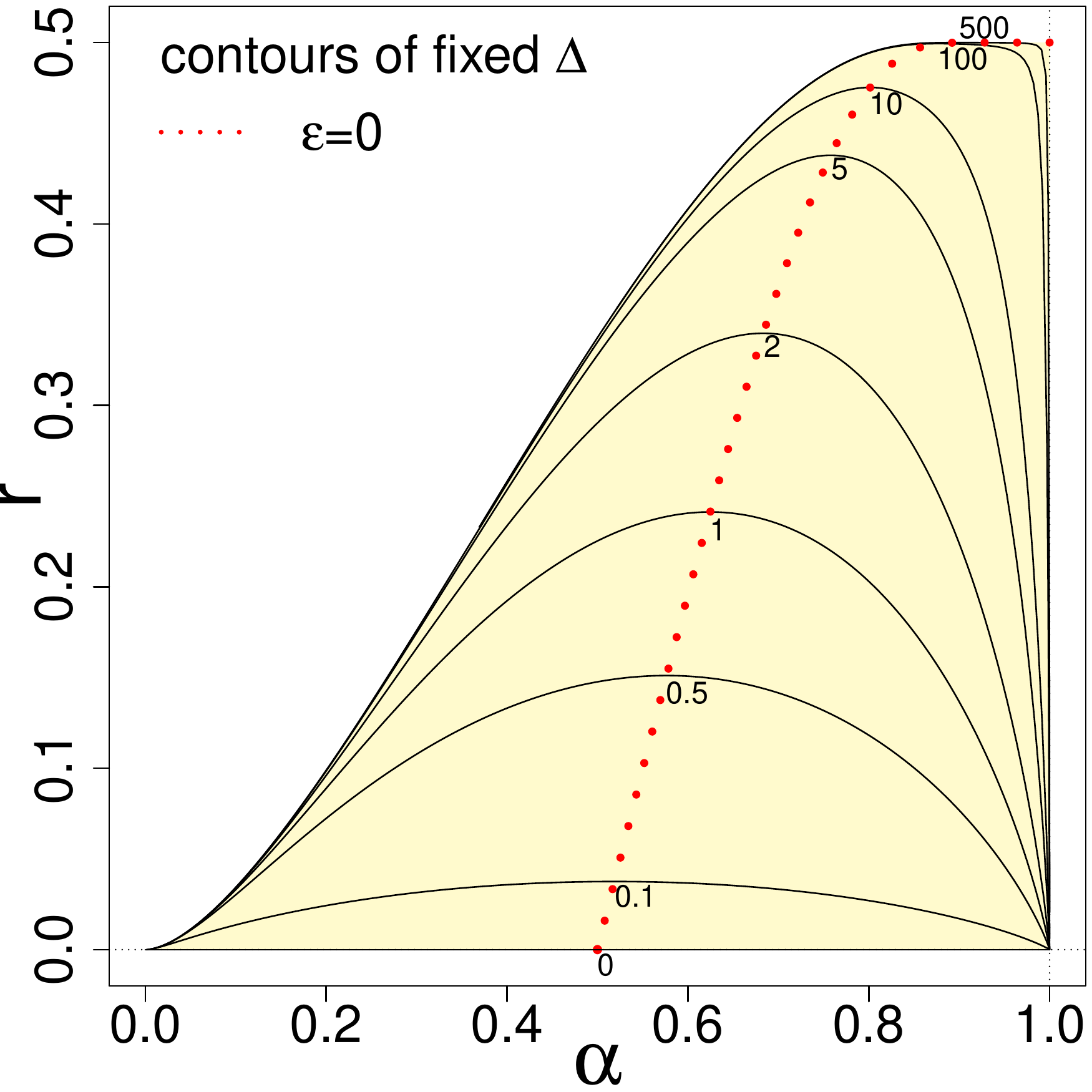}
    \caption{\footnotesize Contour lines of fixed $\Delta$. The red dots represent the maximal values of $r=N/T$ at given $\Delta$ and correspond to the $\epsilon=0$ line. The quantity $\Delta$ is the susceptibility of the portfolio weights to small shifts in the returns, at the same time it is inversely proportional to the estimated in-the-sample ES.}
    \label{fig:Delta}
\end{minipage}
  \qquad
  \begin{minipage}[t][][t]{65mm}
    \includegraphics[width=65mm]{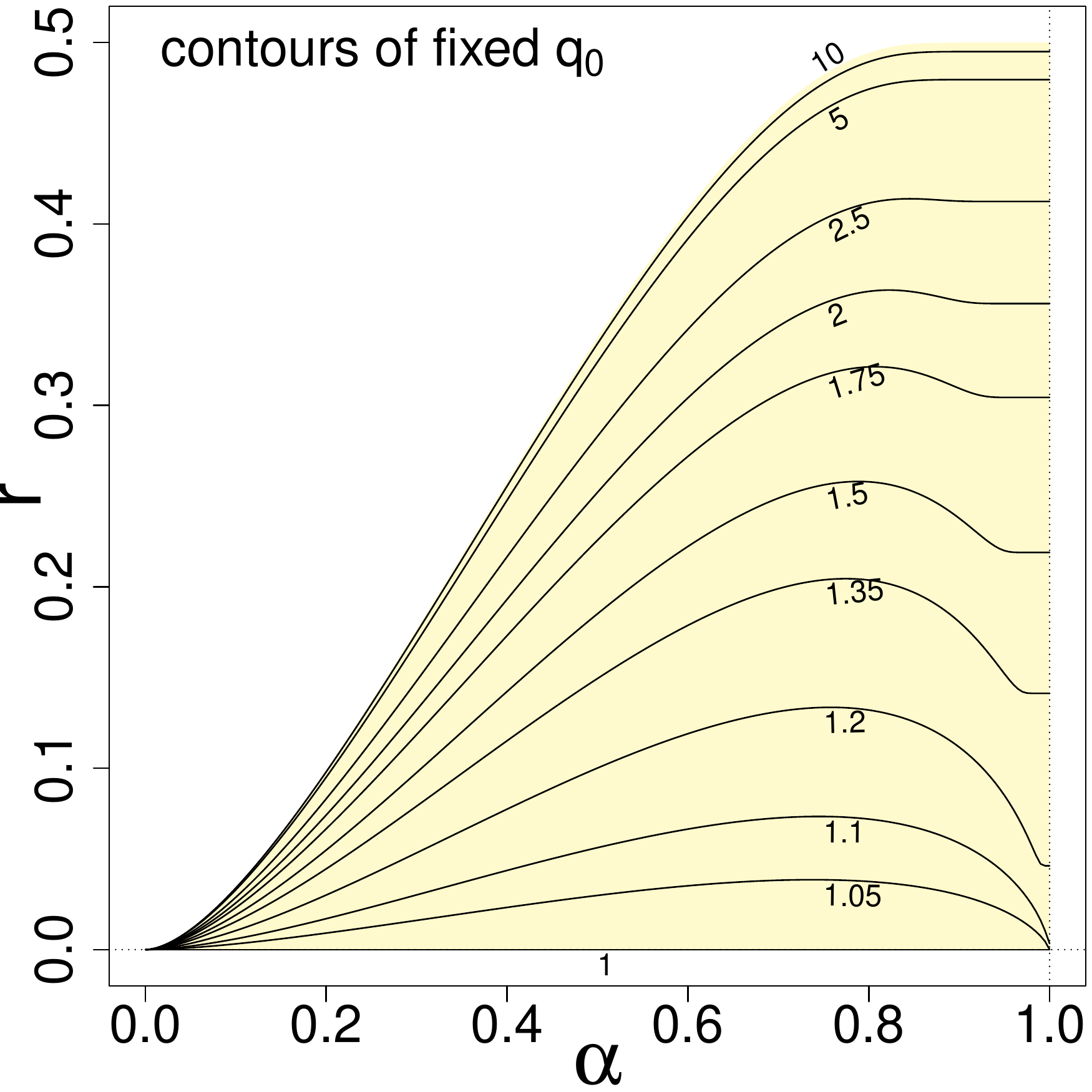}
    \caption{\footnotesize Contour lines of fixed $q_0$. These curves are also the contour lines for the relative error for the out-of-sample estimate of ES.}
    \label{fig:q0}
  \end{minipage}
\end{figure}

Now let us include the third equation~(\ref{equationW}) that determines $\Delta$ in terms of the control parameters and of the two ratios we discussed above, thereby allowing for a complete solution for all three order parameters separately. Fig.~\ref{fig:Delta} displays the contour lines of $\Delta$. These lines more or less follow the phase boundary, until at a point they bend over and fall off towards the point $r=0, \alpha=1$. For higher and higher values of $\Delta$ these contour lines run closer and closer to the phase boundary before they bend over, whereafter they lean tighter and tighter against the vertical line at $\alpha=1$. Note that the contour lines of $\Delta$ never leave the feasible region. What does this behaviour tell us? We have to remember that $\Delta$ appers in two roles: It is inversely proportional to the in-the-sample estimate for ES, Eq.~(\ref{equationESCost}), and it is also the susceptibility of the sample averaged portfolio weights, Eq.~(\ref{equationDeltaMeaning}). The divergence of $\Delta$ means that the in-the-sample average of ES (and also its estimation error) vanish on the phase boundary, precisely at the place where the out-of-sample estimate diverges. It is obvious that the in-the-sample estimation error is always smaller than the out-of-sample one. However, we learn more here. The fact that the ratio $\frac{\Delta}{\sqrt{q_0}}$ is finite when crossing the phase boundary is equivalent to saying that the in-the-sample and out-of-sample estimation errors are inversely proportional to each other at the critical point: the in-the-sample estimate seems to be the most encouraging where it becomes the most misleading. We observed a similar behaviour also in the case of the variance as risk measure \cite{pafka2003Noisy}.

Let us turn to the other two order parameters now. In Fig.~\ref{fig:q0} we show the contour map of $q_0$, the measure of the relative estimation error of ES. As can be seen, the contour lines of $q_0$ also bend over, but in contrast to the $\Delta$ lines, they do not fall down to zero, but after another bend go to some finite value at $\alpha=1$. However, for reasonably small relative errors (corresponding to the lowest curves), this limiting value is very small, implying very large $T$ values.

\begin{figure}[h]
  \centering
    \includegraphics[width=65mm]{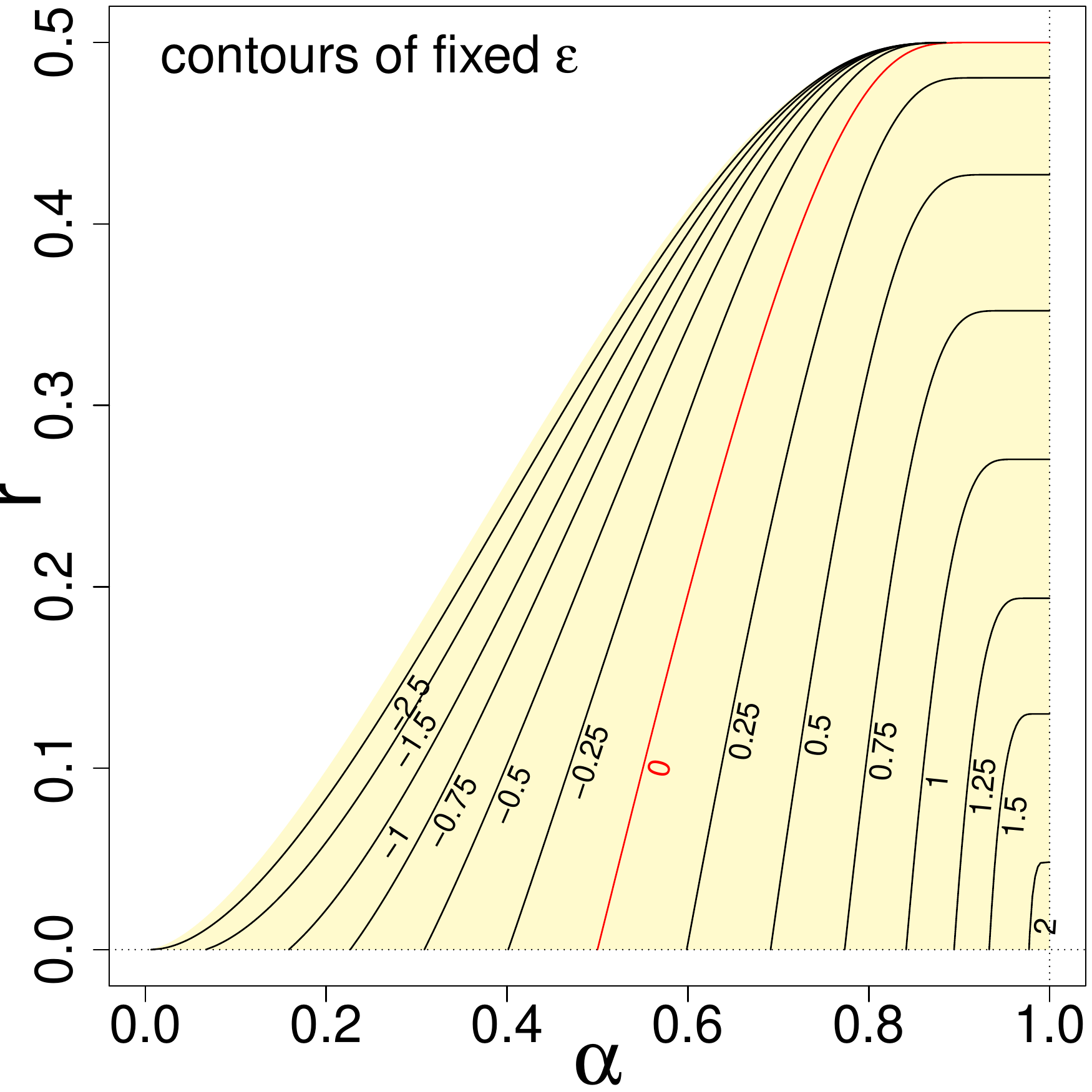}
    \caption{\footnotesize Contour map of $\epsilon$, the VaR of the portfolio optimized under ES.}
    \label{fig:epsilon}
\end{figure}

Finally, in Fig.~\ref{fig:epsilon} the contour map of $\epsilon$ is exhibited. As we have already mentioned, $\epsilon$ is the VaR of the portfolio optimized under ES, and is certainly different from the VaR of a portfolio whose weights are optimized under VaR itself.

\section{ Results for the historical estimates}

\begin{figure}[h]
  \centering
  \begin{minipage}[t][][t]{65mm}
    \includegraphics[width=65mm]{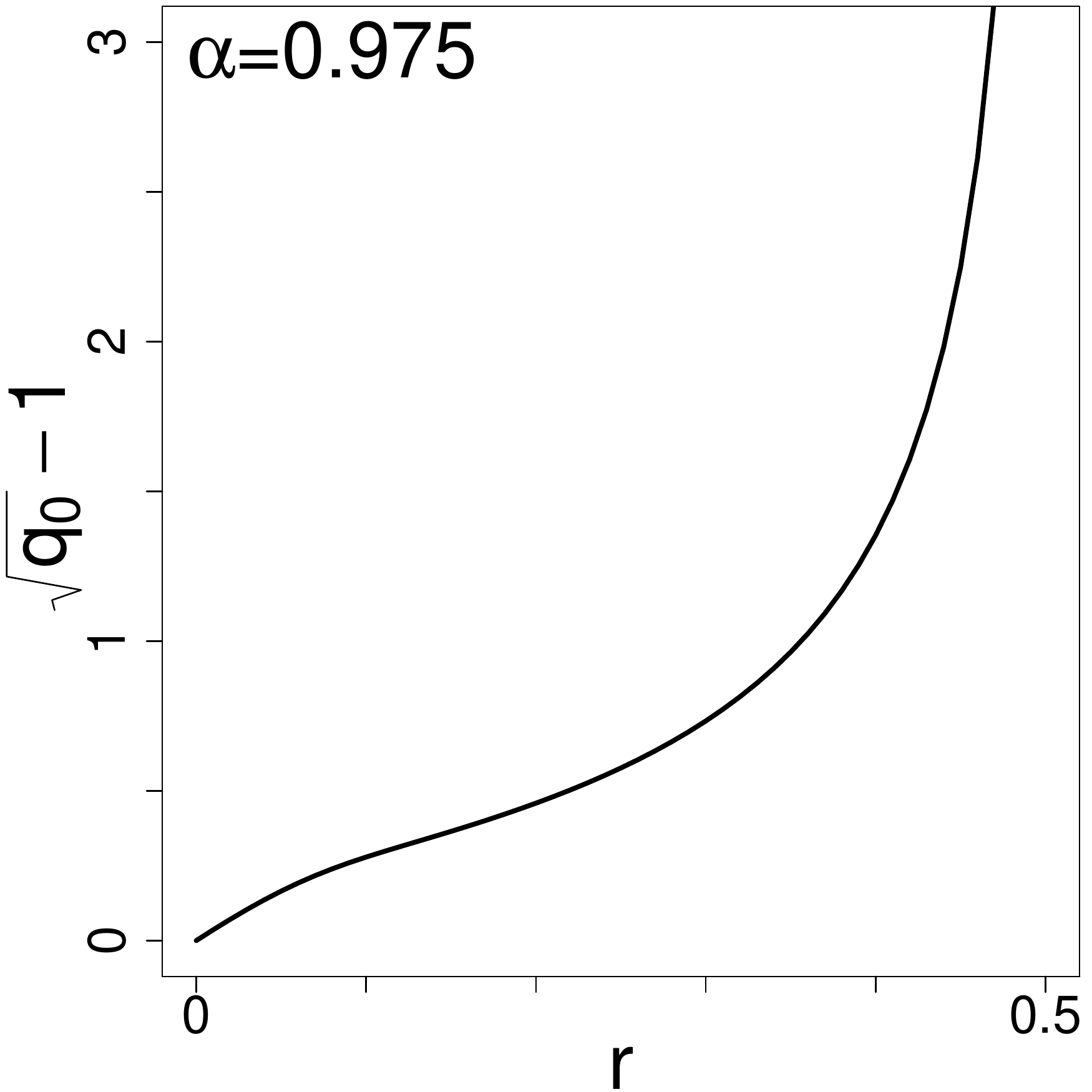}
    \caption{\footnotesize The relative estimation error $\sqrt{q_0} - 1$ of ES as function of $N/T$ at $\alpha$=97.5\%.}
    \label{fig:alp975_q0}
  \end{minipage}
  \qquad
 \begin{minipage}[t][][t]{65mm}
    \includegraphics[width=65mm]{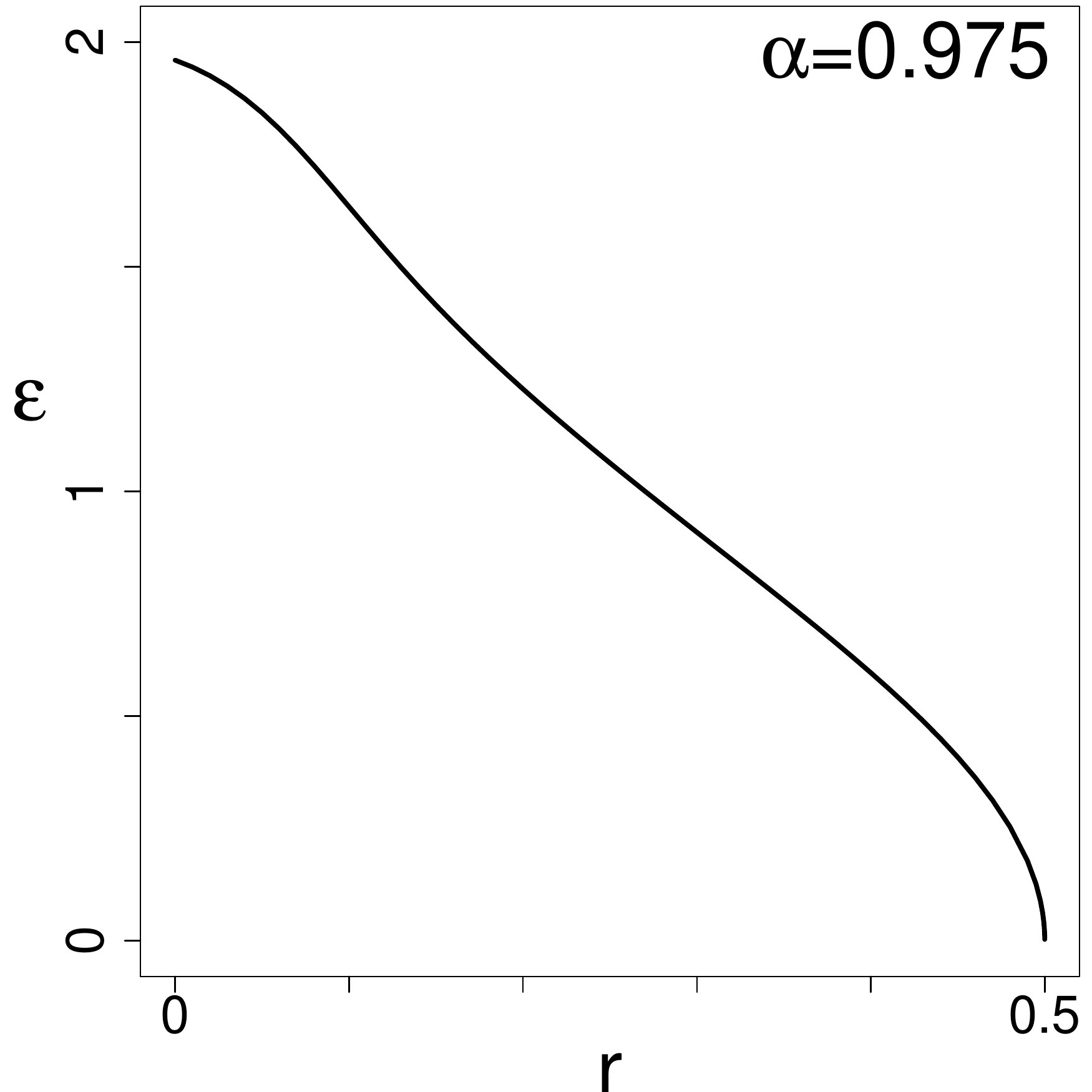}
    \caption{\footnotesize The VaR $\epsilon$ of the ES-optimized portfolio as function of $N/T$ at $\alpha$=97.5\%. The value of $\epsilon$ is monotonically decreasing with increasing $N/T$, and tends to zero as $N/T$ approaches the value corresponding to the phase boundary (very close to 0.5 for $\alpha$ = 0.975).}
    \label{fig:alp975_epsilon}
  \end{minipage}
\end{figure}

\begin{figure}[h]
  \centering
 \begin{minipage}[t][][t]{65mm}
   \includegraphics[width=65mm]{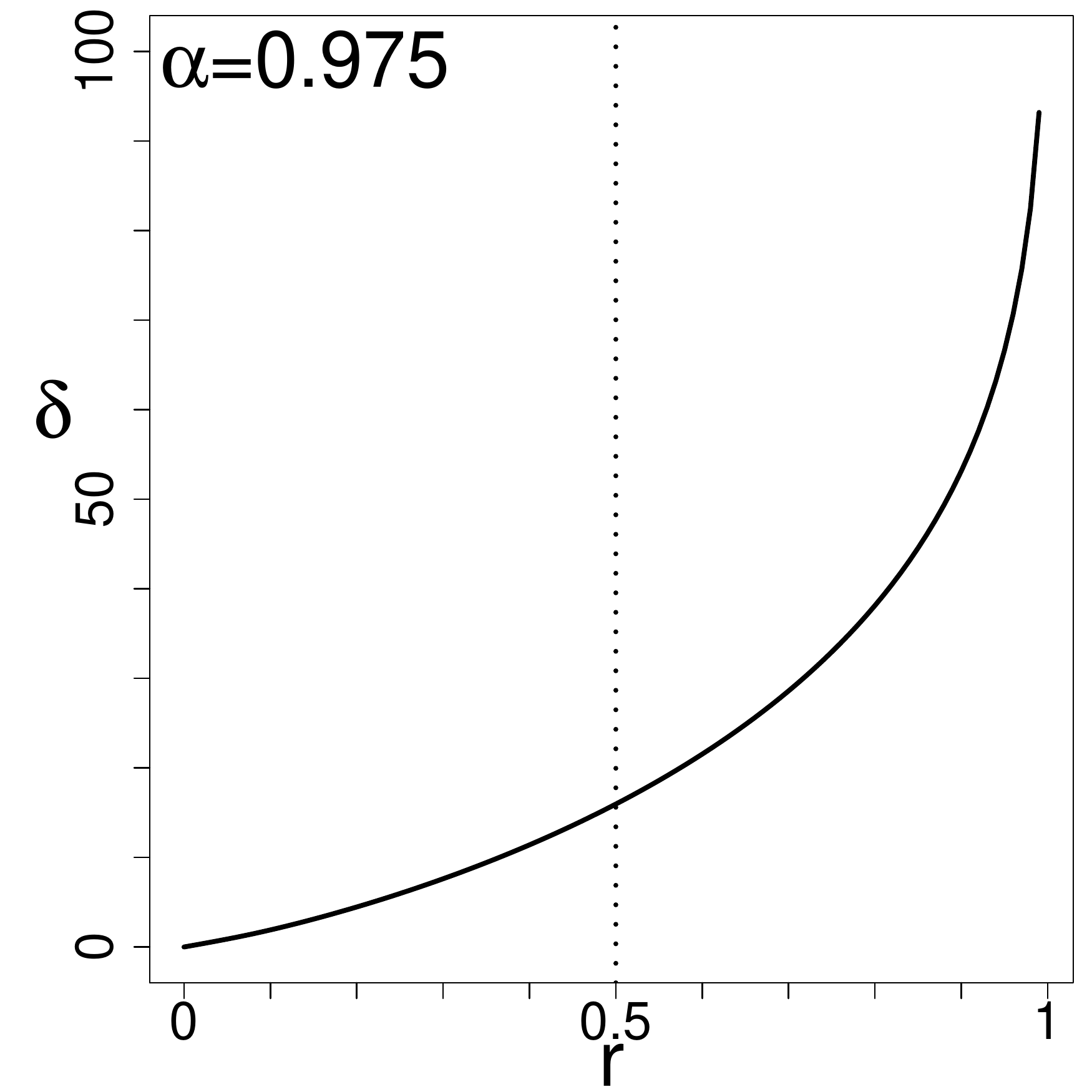}
    \caption{\footnotesize The quantity $\frac{\Delta }{\sqrt{q_0}} =\delta$ measuring the sensitivity of the relative error of the estimated ES, as function of $N/T$ at $\alpha$=97.5\%. The vertical dotted line corresponds to the critical value of $N/T$. The ratio $\frac{\Delta }{\sqrt{q_0} }=\delta$ does not diverge here.}
    \label{fig:alp975_smalldelta}
  \end{minipage}
  \qquad
  \begin{minipage}[t][][t]{65mm}
    \includegraphics[width=65mm]{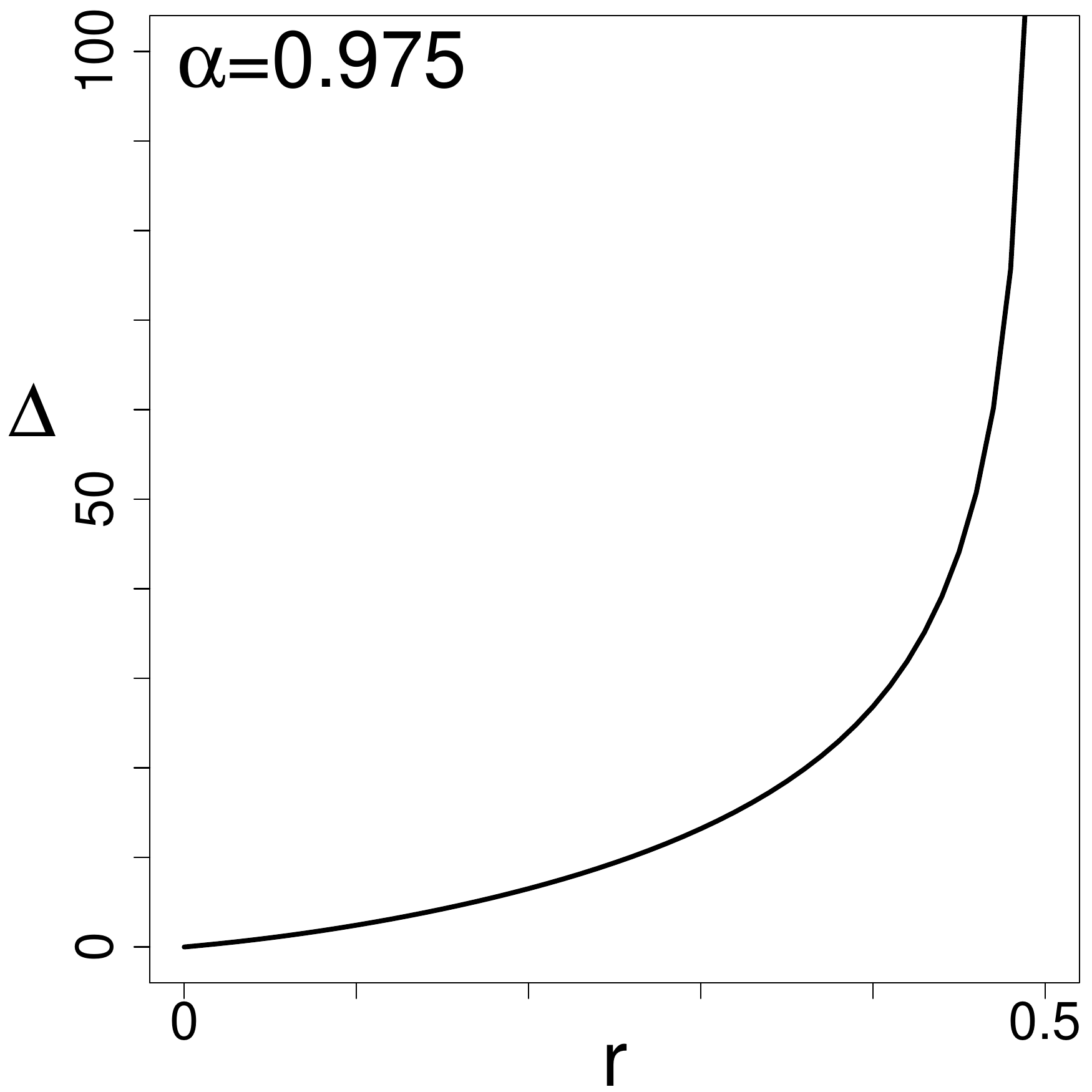}
    \caption{\footnotesize The quantity $\Delta$, the measure of the sensitivity of the portfolio weights to small changes in the returns, as function of $N/T$ at $\alpha=97.5\%$}
    \label{fig:alp975_Delta}
  \end{minipage}
\end{figure}

We are now in a position to draw the consequences of the findings above. In the previous section we constructed the contour maps of the quantities that characterize the estimation error problem of VaR. These maps cover the whole area below the phase boundary where the optimization of ES can be carried out. From a practical point of view the most important region is the vicinity of the $\alpha=1$ line. Let us therefore focus on the line $\alpha=0.975$ advocated by the regulation. The four quantities $\sqrt{q_0}-1$, $\Delta$, $\frac{\Delta}{\sqrt{q_0}}$ and $\epsilon$ as functions of $r$ along the $\alpha=0.975$ line are shown in Figs.~\ref{fig:alp975_q0},\ref{fig:alp975_epsilon},\ref{fig:alp975_smalldelta} and~\ref{fig:alp975_Delta}, respectively.

We can see that $\sqrt{q_0}$ and $\Delta$ are monotonically increasing with $r$. We have learned that $\sqrt{q_0}-1$ is the relative estimation error of ES. According to Fig.~\ref{fig:alp975_q0}, this relative error is small only as long as $N/T$ is small, that is the sample size is large compared to $N$. With $r$ increasing, the relative estimation error quickly becomes very large. Several numerical examples are given in Table~\ref{tab:historical_r}.
\begin{table}[H]
\begin{tabular}{ | r || r |  r | r |  r |  r |  r |  r |  r |  r  | r | r |  r | }
\hline
   estimation & \multicolumn{12}{|c|}{$\alpha$} \\ \cline{2-13}
   error $\downarrow$ &
    0.7 & 0.8 & 0.9 & 0.91 & 0.92 & 0.93 & 0.94 & 0.95 & 0.96 & 0.97 & 0.975 &  0.98 \\ \hline\hline
5\%  &  26 & 27 & 33 & 35 & 37 & 39 & 43 & 47 & 53 & 64 & 72 & 83  \\ \hline 
10\%  & 14 & 14 & 17 & 18 & 19 & 20 & 21 & 24 & 27 & 31 & 35 & 40 \\ \hline
 15\%  & 10 & 10 & 12 & 12 & 13 & 13 & 14 & 16 & 18 & 20 & 22 & 25  \\ \hline
 20\%  & 8 & 8 & 9 & 9 & 10 & 10 & 11 & 12 & 13 & 15 & 16 & 17  \\ \hline
25\%  & 6 & 6 & 7 & 8 & 8 & 8 & 9 & 9 & 10 & 11 & 12 & 12 \\  \hline
50\% &  4 &  4 & 4 & 4 & 4 & 4 & 5 & 5 & 5 & 5 & 5 & 5\\ \hline
 \hline
 \end{tabular}
\caption{\footnotesize The table shows the (rounded) values of $T/N$ that are needed to have a given estimation error for different values of the confidence level $\alpha$. Even an estimation error of $25\%$ requires samples that are $12$ times larger than the number of items in the portfolios at the confidence level $\alpha=0.975$ proposed by regulation.}
\label{tab:historical_r}
\end{table}

This table demonstrates that in order to have a 10\% or 5\% relative error in the estimated Expected Shortfall of a moderatetly large portfolio of, say, N=100 stocks at $\alpha=0.975$ we must have time series of length $T=3500$ resp. $T=7200$. These figures are totally unrealistic: they would correspond to 14 resp. 28.8 years even if the time step were taken as a day (rather than a week or month) . 

The behaviour of $\Delta$, the measure of the sensitivity of the optimal weights to small changes in the returns, is similarly discouraging: it grows very fast and diverges at the phase boundary. As for the susceptibility $\frac{\Delta}{\sqrt{q_0}}$ that measures the sensitivity of the ES estimate, it also increases fast with $N/T$, though it remains finite at the phase boundary.

In contrast to the above, $\epsilon$, the VaR of the ES-optimized portfolio, is decreasing with increasing $r$. This is in accord with the behaviour of the in-the-sample estimate of ES itself (proportional to the reciprocal of $\Delta$) that vanishes at the phase boundary.

The vanishing of the in-the-sample ES and VaR at the phase boundary can be understood by considering that as we approach the phase boundary the apparent arbitrage effect is dominating the optimization more and more, so the probability density of the optimal portfolio (not the density of the weights, but of the profit and loss distribution) shifts to the left (remember that by convention loss is regarded as positive and gain negative). As a consequence, ES and VaR corresponding to a fixed $\alpha$ must decrease monotonically.

 \section{Contour lines of the error of parametric ES estimates}
 So far we have considered historical estimates of ES and seen that the estimation error is very large for any reasonable set of parameters (portfolio size, confidence  level, sample size), or conversely, that the time series necessary to produce an acceptable estimation error are extremely long. We may expect that parametric estimates fare rather better, and this is what we are going to show in this Section. 

To make the difference between the two approaches clear, we note that although in the preceeding sections we used a Gaussian distribution to generate the data for returns, during the course of optimization we pretended as if we had not known this fact, and treated those data 
as if they had been observed in the market. In contrast, in this section we will assume that the data follow a Gaussian distribution, but we do not know its parameters (mean and variance). Actually, this problem has been considered in \cite{varga2008TheInstability}, but the focus in that paper was on the problem of instability again and the degree of estimation error inside the feasible region was not investigated. The solution was obtained by the method of replicas, and followed by and large the same lines as the treatment of the historical estimate, only it was somewhat simpler. Having recapitulated the key points of the replica method in the context of the historical estimate, we feel we do not need to go into any details now, so we just refer the reader to the paper \cite{varga2008TheInstability}, and pick up the thread at the formula~(\ref{EqFirstOrder2}) there. (Note that the quantity $q_0$ was called $q_0^2$ in \cite{varga2008TheInstability}.) This formula gives the average over the samples of the square of the estimation error $\sqrt{q_0}$ as
\be
\label{Eqq0avg}
 {q_0} = \frac{\phi(\alpha)}{(1-r) \phi(\alpha)-r} = \frac{r_c(\alpha)}{r_c(\alpha)-r} .
\ee
 
 where
 
\be
\phi(\alpha) =  \frac{e^{-\frac12 \left( \Phi^{-1}(\alpha) \right)^2}}{(1-\alpha) \sqrt{2\pi}}  ,
\ee

 $\alpha$ is the confidence level and $\Phi^{-1}$ the inverse of the cumulative standard normal distribution, as before.

As for $r_c$, it is the critical value of $r=N/T$ at which  the average estimation error diverges
\be
  r_c(\alpha) = \frac{\phi^2(\alpha)}{1+\phi^2(\alpha)} \quad .
\ee

General theoretical considerations \cite{mezard1987Spin} supported by numerical evidence suggest that in the limit of large $N$ the distribution of $q_0$ over the samples is sharp, so we may take the liberty of regarding $q_0$ as a given number rather than a random variable.

\begin{figure}[h]
  \centering
    \includegraphics[width=65mm]{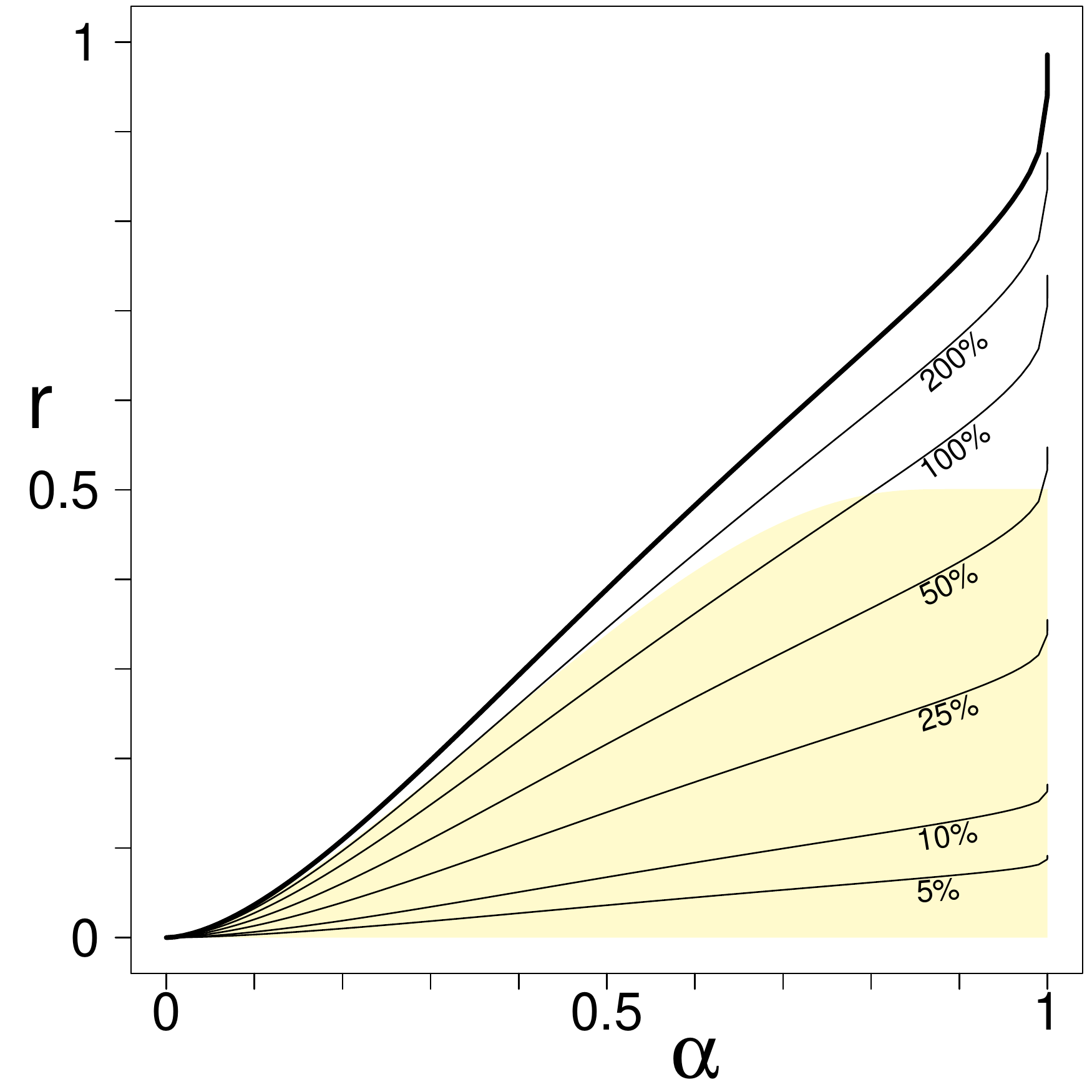}
   \caption{\footnotesize The contour map of the error of the parametric estimates for ES.}
    \label{fig:rc_parametric}
\end{figure}

We can see from~(\ref{Eqq0avg}) that $q_0$ diverges when $r$ goes to $r_c$ from below: at this point the parametric estimate loses its meaning. The curve $r=r_c$ is the phase boundary for the parametric estimates for ES. It is the uppermost curve in Fig.~\ref{fig:rc_parametric} .

To obtain the contour lines we invert the formula~(\ref{Eqq0avg}) and express $r$ as:
\be
  r = \frac{q_0 -1}{q_0} r_c(\alpha) .
\ee
As can be seen, the lines belonging to a given value $q_0$ of the estimation error are simply scaled down from the critical line. As $q_0$ is larger than or equal to 1 by definition, the factor in front of $r_c(\alpha)$ varies between zero (corresponding to $q_0=1$, that is to an infinitely long observation time, $N/T=0$) and one (corresponding to $q_0=\infty$ on the phase boundary). 

A few of these curves are shown in Fig.~\ref{fig:rc_parametric}.

It is now easy to work out the sample size $T$ necessary for a given relative error and a given size $N$ of the portfolio. Let us consider the parametric estimate of ES for e.g. a portfolio of $N=100$ different securities, and stipulate a relative error of 10\%, i.e. $q_0=1.1$. Let us assume furthermore that the confidence level is $\alpha=0.975$ as envisaged in regulation. The critical value $r_c$ at this 
$\alpha$ is about 0.9, so $r$ works out to be about 0.156. For a portfolio size $N=100$ this means that the length of the necessary time series to ensure the 10\% error is 682 time steps (days or weeks, depending on the observation frequency of the portfolio manager). This is a very large number, though much less than the 3500 steps needed for the same precision in the historical estimate. Several further numerical examples are given in Table~\ref{tab:parametric_r}.
\begin{table}[h]
\begin{tabular}{ | r || r |  r | r |  r |  r |  r |  r |  r |  r  | r | r |  r | }
\hline
   estimation & \multicolumn{12}{|c|}{$\alpha$} \\ \cline{2-13}
   error $\downarrow$ &
    0.7 & 0.8 & 0.9 & 0.91 & 0.92 & 0.93 & 0.94 & 0.95 & 0.96 & 0.97 & 0.975 &  0.98 \\ \hline\hline
5\%  &  19 & 16 & 14 & 14 & 14 & 14 & 13 & 13 & 13 & 13 & 13 & 13  \\ \hline 
10\%  & 10 & 9 & 8 & 8 & 7 & 7 & 7 & 7 & 7 & 7 & 7 & 7 \\ \hline
 15\%  & 7 & 6 & 5 & 5 & 5 & 5 & 5 & 5 & 5 & 5 & 5 & 5  \\ \hline
 20\%  & 6 & 5 & 4 & 4 & 4 & 4 & 4 & 4 & 4 & 4 & 4 & 4  \\ \hline
25\%  & 5 & 4 & 4 & 4 & 4 & 4 & 3 & 3 & 3 & 3 & 3 & 3 \\  \hline
50\% &  3 &  3 & 2 & 2 & 2 & 2 & 2 & 2 & 2 & 2 & 2 & 2\\ \hline
 \hline
 \end{tabular}
\caption{\footnotesize The table shows the rounded values of $T/N$ that are needed to have a given estimation error for different values of the confidence level $\alpha$ used in the calculation of the parametric estimate for Expected Shortfall. }
\label{tab:parametric_r}
\end{table}

If we are a little more demanding and prescribe an estimation error of 5\%, these numbers work out to be about $T=1272$, resp. 7200 for the parametric, resp. historical estimate.

Although the contour map of parametric VaR is not a subject of this paper, from \cite{varga2008TheInstability} we know that the difference between the ES and VaR level curves must be negligible in the region of $\alpha$'s in the vicinity of 1, thus the data requirements of parametric VaR estimation would be as absurd as in the case of ES.

We can see that the parametric estimates are less data demanding than the historical estimates, as expected, but they are still in a range which is totally beyond any practically achievable sample size.

\section{Remarks on possible extensions: correlations and inhomogeneous portfolios, fat tailed distributions, and regularization}

We have made a number of simplifying assumptions in this study: we assumed that the fluctuations of the underlying risk factors were i.i.d. normal, disregarded all the possible constraints except the budget constraint, and considered the special limit $N,T\to\infty$ with $N/T$ finite. One may wonder how tightly these assumptions are linked to the disappointing results for the estimation error, and whether any of them can be relaxed. 

Let us first consider the question of identical distribution and independence. As shown in  \cite{varga2008TheInstability} in the case of the parametric estimates for VaR and ES, but equally true for the historical estimates, Gaussian fluctuations with an arbitrary (but invertible) covariance matrix can simply be accommodated in the replica formalism at the expense of some additional effort and even more complicated formulae. Likewise, a constraint on the expected return of the portfolio can easily be included, adding one more Lagrange multiplyer to the problem. All these features leave the essence of our message intact, in fact, they demand even larger samples for the same level of estimation error than in the simplified problem we analyzed above. 

\begin{figure}[h]
  \centering
    \includegraphics[width=65mm]{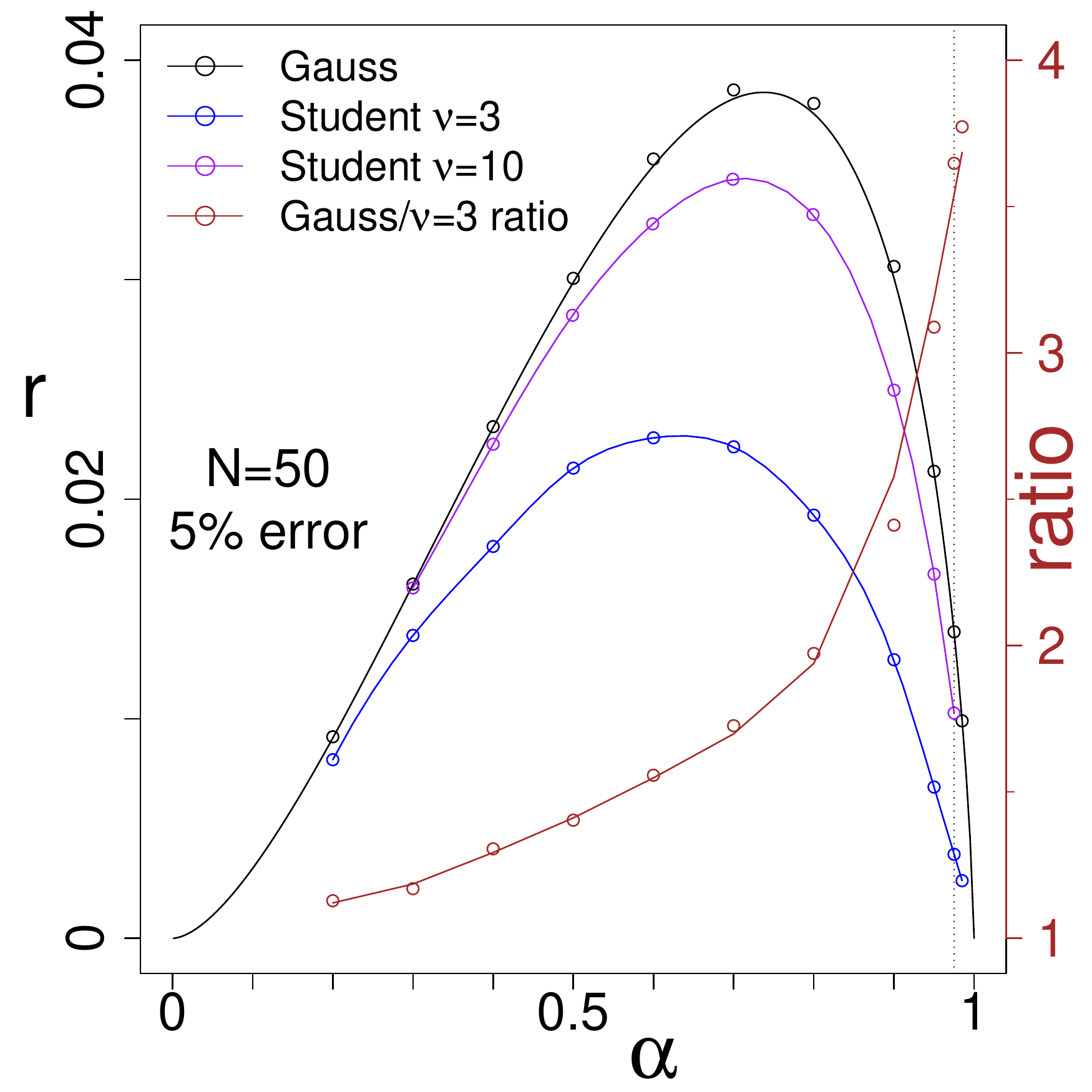}
    \caption{\footnotesize Estimation error $\sqrt{q_0}-1 = 0.05$ contour line obtained from numerical simulations for a portfolio size $N=50$ for Gaussian (small black circles), Student $\nu=3$ (blue line and small blue circles) and Student $\nu=10$ (purple line and purple circles) distributions. For comparison the replica theoretic result is also presented (black line). The brown line shows the ratio of the $N/T$ values corresponding to the same $\alpha$ for the Gaussian and the Student $\nu=3$.}
    \label{fig:delta_student}
\end{figure}

The Gaussian character of the underlying fluctuations is, however, an essential limitation: the replica formalism cannot cope with non-Gaussian underlying fluctuations, whereas they are a general feature of real markets. In order to study the effect of fat tails, we had to resort to numerical simulations to solve the linear programming problem in (\ref{eqCostFunction}). As could be expected, fat tails make the estimation errors even larger than the Gaussian fluctuations. An example is shown in Fig.~\ref{fig:delta_student} where we show simulation results for the level curve corresponding to $\sqrt{q_0}-1 = 0.05$ that is a 5\% error in the  the out-of-sample historical estimate of ES for the Gaussian case along with the same curves for two Student distributions with $\nu=3$, resp. $\nu=10$ degrees of freedom. 

This figure needs a few comments. The continuous black line comes from the analytical replica theoretic calculation, the small black circles are the results of simulations (numerical solutions of the linear programming problem) at the corresponding values of the control parameters. Notice that the simulation results essentially fall on the analytical curve already at this relatively small value of $N=50$. This is a general experience in the interior of the feasible region: simulation of relatively modest size portfolios with $N$'s in the range 50 to a few hundred reproduce the analytical results (corresponding to the limit $N\to\infty$) quite well, provided the numerical results are averaged over a sufficiently large number (often 500 and above) of samples. In the immediate vicinity of the phase boundary, however, convergence slows down considerably, and the portfolio size and the number of samples required for a precise numerical result quickly grow out of the practically achievable range.

The simulations for i.i.d. Student distributed returns with $\nu=3$ resp. $\nu=10$ degrees of freedom, and with the same $N=50$ and 5\% error as in the Gaussian case, produce the contour lines shown in blue ($\nu=3$) and purple ($\nu=10$), respectively. (The continuous blue and purple lines are just guides to the eye, the measured data are shown by the small circles.) As expected, the contour lines corresponding to these fat tailed distributions lie below the Gaussian curve, which means that to have the same estimation error for a given $N$ one needs even larger samples than in the Gaussian case. The $\nu=10$ Student curve is much closer to the Gaussian one than the $\nu=3$ curve, which is how it should be: for $\nu\to\infty$ the Student distribution goes over into the Gaussian. The difference between the Gaussian and the $\nu=3$ Student curves increases fast as we approach $\alpha=1$: the ratio of the two are shown by the brown line. The vertical dotted line marks the regulatory value of $\alpha=0.975$. The ratio between the Gaussian and the $\nu=3$ Student values is about 3.7 at this $\alpha$.
This means one has to have almost four times larger samples for such a fat tailed distribution than for a Gaussian. This is plausible: we are dealing with a risk measure focusing on the fluctuations at the far tail where the difference between a narrow distribution and a fat tailed one is the largest. The difference between the Gaussian and the $\nu=3$ curves is certainly not small, but the data requirement is so unrealistic already in the Gaussian case that the additional demand for fat tailed distributions is almost immaterial.

\begin{figure}[h]
  \centering
    \includegraphics[width=65mm]{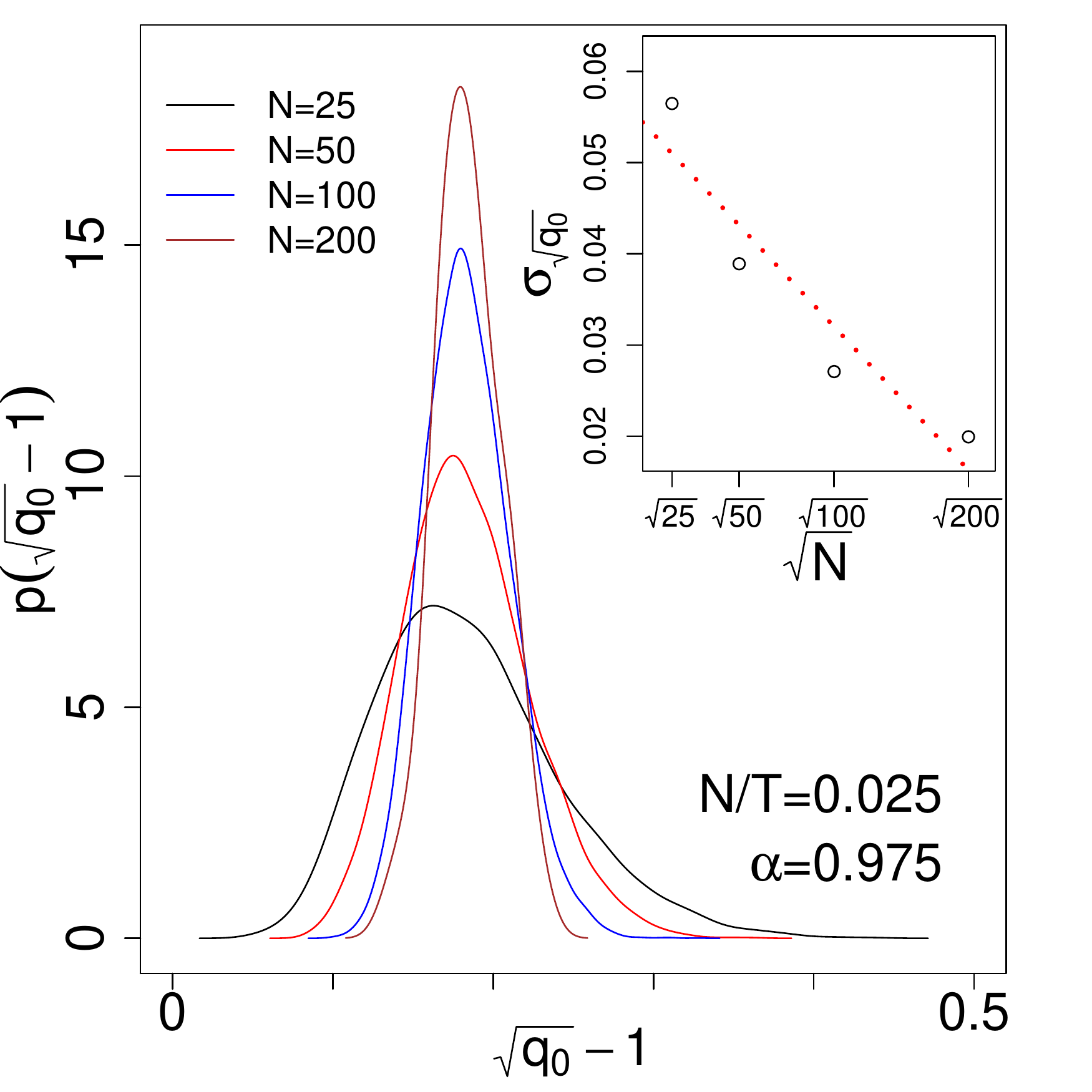}
    \caption{\footnotesize Distribution of the estimation error over the samples from numerical simulations at $N/T=0.025$ for $N=25,50,100,200$ at the confidence level $\alpha=97.5\%$. The curves were obtained by averaging over 5000, 5000, 15000, and 100 samples. The tendency of the distribution becoming sharper and sharper as $N,T\to\infty$ is clear. Inset: the dependence of the extracted width of the estimation error distribution. The width approaches 0 in the limit $N,T\to\infty$.}
    \label{fig:num_q0_Ndep}
\end{figure}

While the replica method enables us to calculate the expectation value of the relative error of ES and the distribution of optimal portfolio weights averaged over the random Gaussian samples, it does not provide information about how strongly these quantities fluctuate from sample to sample. (This is not a limitation in principle: pushing the saddle-point calculation in the background of the replica method one step beyond leading order, one could derive the width of the distribution of estimation error. Such a calculation would demand a very serious effort and is beyond the scope of the present work.)  Instead of trying to derive the width of the distribution analytically, we have resorted to numerical simulations again. We have found that at a safe distance from the phase boundary the distribution of the estimated ES over the samples is becoming more and more concentrated, its width approaching zero in the $N,T\to\infty$ limit. While the position of the peak of the distribution stabilizes fairly fast, the convergence of the width is rather slow. An illustration is given in Fig.~\ref{fig:num_q0_Ndep}. In the vicinity of the phase boundary, however, the average estimation error grows beyond any bound, and there its fluctuations depend on the order of limits: if we go to the phase boundary while keeping $N,T$ finite, the width of the distribution blows up, in the opposite limit the distribution evventually shrinks into a Dirac delta. This behaviour is in accord with what one expects to find at a phase transition.

Another essential limitation of our analysis is the omission of the possible constraints on the portfolio weights. These constraints may impose limits on short selling, on different groups of assets (corresponding to industrial sectors, geographical regions, etc.). The constraints can (and typically do) confine the domain where the optimum is sought to a finite volume in the space of portfolio weights. Any such constraint will act as a regularizer, and will prevent the phase transition to the infeasible region to take place. Regularizers can be built into the replica approach, as has been demonstrated in \cite{caccioli2013Optimal,caccioli2014Lp}. However, the various possible limit systems or regularizers introduce very different modifications to the optimization problem, and we thought their inclusion and detailed analysis would lead too far from the main thrust of the present paper (and would at least double its length), therefore we decided to leave this important aspect of the problem to a subsequent publication.

\section{Conclusions}

Expanding on a short, and very preliminary, account of this work \cite{kondor2015Contour}, in this paper we considered the problem of optimal portfolio selection under the risk measure Expected Shortfall in the simplest concievable setting: we assumed that the portfolio is a linear combination of assests with i.i.d. standard normal distributed returns and we aimed at the global minimum risk portfolio, omitting any constraints on the weights except the budget constraint. Thus the underlying process did not have any structure at all, it was just pure noise. The basic question we asked was: for a given number $N$ of assets in the portfolio, how large the size $T$ of the statistical samples must be, in order for the optimization to return the correct optimum (with no structure and the optimal portfolio weights all equal to one another) within a prescribed estimation error. The answer we obtained in the case of historical estimation was very discouraging: optimizing over finite samples produced a typically broad weight distribution, imaginary correlation structures and a mirage of arbitrage. In order to obtain the correct answer within an acceptable error, we would have needed exceedingly large samples, far beyond any length of observation times achievable in practice. Qualitatively, this is just what one would have expected, but our results help articulate this conclusion in a precise, quantitative manner.

Remarkably, though the results for parametric estimates turned out to be much more favourable than those for the historical estimation, the necessary sample sizes remained way above anything that could be regarded as realistic.

Our conclusion is thus in agreement with a number of other authors who consider the task of optimizing large portfolios in its original (unfiltered, unregularized) form as hopeless. To paraphrase the first sentence of the Abstract in \cite{ledoit2004Honey}: The central message of this paper is that nobody should be using Expected Shortfall for the purpose of portfolio optimization. One might raise the objection that ES is meant to be a diagnostic tool to measure the risk of a portfolio, not an aid to decision making concerning its optimal construction. This may well be the case, but the career of VaR \cite{Grootveld2004Upgrading} suggests that institutions will inevitably be driven to use the new regulatory market risk measure beyond its purported scope: a binding constraint easily assumes the role of an objective function. Besides, risk is hard not only to optimize, but as the findings of Danielsson and Zhou \cite{danielsson2015Why} clearly demonstrate, also to measure.

The method we applied to reach our quantitative results is, to the best of our knowledge, the only known analytical approach to the optimization of ES and led us to a closed set of equations for the  relative error of the out-of sample estimate, the sensitivity to small changes in the returns, and for the VaR of the ES-optimized portfolio. These equations were solved numerically; in some special cases, including that of the interesting minimax risk measure, by hand. The domain of applicability of our method is not limited to the trivial case of i.i.d random variables and we plan to use it to models where the underlying stochastic process has a structure (inhomogeneous portfolios with non-zero covariances and returns). It will be interesting to see how large the size of samples must be in order to recover such a structure with a tolerable error.

In the present work, we deliberately left out any regularization or other dimensional reduction method from consideration. The application of these methods is mandatory in a high-dimensional setting such as the present one. Their omission here is motivated partly by trying to present, in its simplest form, a method that may be unfamiliar for most readers, together with the nontrivial analytic results it leads to, but also by trying to keep the length of the paper within reasonable bounds. There is a more serious consideration, however. Regularization, dimensional reduction, limits on groups of weights, or any other method designed to restrain the wild fluctuations of estimates, impose a certain structure on the acceptable optima thereby necessarily introducing bias. This raises the nontrivial question about the tradeoff between bias and fluctuation. In view of the extremely unstable estimates, a very strong regularization would be needed to stabilize them, so strong indeed as to act as a dominating Bayesian prior, basically supressing the information coming from the empirical samples. Faced with finite samples of real-life market data (instead of our synthetic data), a portfolio manager may decide to entirely disregard the information coming from the market and go for the naive $1/N$ portfolio (as suggested by the result of \cite{DeMiguel2009Optimal}), or act on expert opinion or gut feeling, what may be happening in most cases anyhow \cite{Michaud1989TheMarkowitz}. Or, worse still, she may trust a black-box optimizer package as a source of cleansed market information. It is a worthy endeavour to clarify the precise quantitative conditions under which background knowledge can be incorporated into optimization, without bias crowding out market information completely, and we believe the methods employed in the present paper in an oversimplified setting will prove to be useful in the case of more realistic  market models as well.

\section*{Acknowledgements}
We are obliged to a number of people for useful interaction on this project, including E. Berlinger, I. Csabai, J. Danielsson, B. D\"om\"ot\"or, F. Ill\'es, R. Kondor, M. Marsili and S. Still. FC acknowledges support of the Economic and Social Research Council (ESRC) in funding the Systemic Risk Centre (ES/K002309/1). IK is grateful for the hospitality extended to him at the Computer Science Department of the University College of London during part of the composition of this paper.

\appendix
\renewcommand{\thesection}{Appendix \Alph{section}}
\renewcommand{\thesubsection}{\Alph{section}.\arabic{subsection}}
\renewcommand{\theequation}{\Alph{section}.\arabic{equation}}
\setcounter{equation}{0}

\section{The replica calculation}
\label{sec:appendixA}

In this Appendix we show how one can deduce the cost function Eq.( \ref{free_energy}) from the linear programming problem that solves the optimization of Expected Shortfall. The method is taken over from the theory of disordered systems and goes by the name of the method of replicas. This method was first applied in the portfolio context by Ciliberti et al. \cite{ciliberti2007On} and the derivation has appeared in slightly modified forms in \cite{kondor2010Instability}; it is included here to make this paper self-contained. 

We need to find the minimum of
$$E[\epsilon,\{u_{t}\}]=(1-\alpha) T\epsilon+\sum_{t=1}^T u_{t}$$
under the constraints
$$u_{t}\ge0,\;\;\;   u_{t}+\epsilon+\sum_{i=1}^N x_{i,t} w_i\ge 0\;\;\; \forall t$$ and $$\sum_{i=1}^N w_i =N.$$
The calculation proceeds as follows:
Following the general strategy of statistical physics, we replace the ``sharp'' optimization above by a ``soft'' one via the introduction of a fictitious inverse temperature $\gamma$ and define the canonical partition function (or generating functional) as

\be
Z_\gamma\left[\{x_{i,t}\}\right]=\int_0^\infty \prod_{i=1}^T d u_t\int_{-\infty}^\infty d\epsilon~ \theta\left(u_{t}+\epsilon+\sum_{i=1}^N x_{i,t} w_i\right) e^{-\gamma E[\epsilon,\{u_{t}\}]},
\ee
where $\theta(x)=1$ if $x>0$ and zero otherwise.

The partition function is therefore an integral over all possible configurations of variables that are compatible with the constraints of the problem, where each configuration $\epsilon,\{u_{t}\}$ is weighted by the Boltzmann weight $e^{-\gamma E[\epsilon,\{u_{t}\}]}$. The original optimization problem can be retrieved in the limit $\gamma\to\infty$ where only the minimal value of $E[\epsilon,\{u_{t}\}]$ contributes. From the partition function, the minimum cost (per asset) can be computed in the limit of large $N$ as
\be
\lim_{N\to\infty}\lim_{\gamma\to\infty}-\frac{\log Z_{\gamma}[\{x_{i,t}\}]}{\gamma N}.
\ee
To derive the typical properties of the ensemble we have to average over all possible realizations of returns and compute
\be
\langle\log Z_{\gamma}\left[\{x_{i,t}\}\right]\rangle=\int_{-\infty}^{\infty} \prod_{i=1}^N\prod_{t=1}^T dx_{i,t} P[\{x_{i,t}\}]\log Z_{\gamma}\left[\{x_{i,t}\}\right],
\ee
where $P[\{x_{i,t}\}]$ is the probability density function of returns.
Averaging a logarithm is difficult. The replica trick has been designed to circumvent this difficulty. It is based on the use of the identity
\be
\langle \log Z\rangle = \lim_{n\to 0}\frac{\partial \langle Z^n\rangle}{\partial n}.
\ee
For integer $n$, we can compute $Z^n$ as the partition function of a system composed of $n$ identical and independent replicas of the original systems. An analytical continuation to real values of $n$ will then allow us to perform the limit $n\to0$ and obtain the sought quantity $\langle\log Z_{\gamma}\left[\{x_{i,t}\}\right]\rangle$. The Achilles heel of the method is the analytic continuation from integer to real $n$'s; the uniqueness of the analytic continuation typically cannot be easily proven. In the theory of disordered systems, the results originally obtained via the replica trick were later verified by rigorous mathematical methods \cite{guerra2002TheThermodynamic,guerra2003TheInfinite,talagrand2003Spin}. Such a rigorous proof has not been constructed for the present model. However, given the convexity of our cost function, we believe that the method is bound to lead to the correct answer. Nevertheless, with a rigorous proof lacking, we regard the replica theory as a heuristic method, so we felt necessary to solve the linear programming problem also by direct numerical methods to verify the results obtained via replicas.

The replicated partition function can be computed as\footnote{In the calculation we will not keep track of constant multiplicative factors that do not affect the final result.}
\begin{eqnarray*}
Z_{\gamma}^n &=& \int_{-\infty}^{\infty}\left(\prod_{i=1}^N\prod_{t=1}^T dx_{i,t}\right)\int_{-\infty}^{\infty} \left(\prod_{a=1}^n d\epsilon^a\right)
\int_0^{\infty}\left(\prod_{t=1}^T\prod_{a=1}^n du_{t}^a\right)\int_{-\infty}^{\infty}\left( \prod_{i=1}^N \prod_{a=1}^n d
w_i^a\right)\\
&\times&\int_{-i\infty}^{i\infty}\left(\prod_{a=1}^n d\hat\lambda^a\right)  \int_{0}^{\infty} \left(\prod_{t=1}^T\prod_{a=1}^n d
\mu_{t}^a\right) \int_{-\infty}^{\infty}\left(\prod_{t=1}^T \prod_{a=1}^n d\hat{\mu}_{t}^a \right)\prod_{t=1}^T \prod_{i=1}^N\exp\left\{-\frac{Nx_{i,t}^2}{2}\right\}\\
&\times&\exp\left\{\sum_a \hat\lambda^a
(\sum_i w_i^a-N)\right\}
\prod_{t}\exp\left\{\sum_a i \hat{\mu}_{t}^a\left(u_{t}^a+\epsilon^a+\sum_i x_{i,t}w_i^a-\mu_{t}^a \right)\right\}\\
&\times& \exp\left\{ -\gamma\sum_a(1-\alpha)T \epsilon^a-\gamma\sum_{a,t} u_{t}^a \right\},
\end{eqnarray*}
where we have assumed that 
\be
P[\{x_{i,t}\}]=\prod_{t=1}^T \prod_{i=1}^N\exp\left\{-\frac{Nx_{i,t}^2}{2}\right\},
\ee 
and we have enforced the constraints through the Lagrange multipliers $\hat\lambda^a$, $\mu_t^a$ and $\hat\mu_t^a$.
Averaging over the quenched variables $\{x_{i,t}\}$ and introducing the overlap matrix
$Q_{a,b}=\frac{1}{N}\sum_i w_i^a w_i^b$  and its conjugate $\hat Q_{a,b}$ one obtains

\begin{eqnarray*}
Z_{\gamma}^n&=& \int_{-i\infty}^{i\infty}\left(\prod_{a=1}^n\prod_{b=1}^n dQ_{a,b}d\hat Q_{a,b}\right)\int_{-\infty}^{\infty} \left(\prod_{a=1}^n d\epsilon^a\right)
\int_0^{\infty}\left(\prod_{t=1}^T\prod_{a=1}^n du_{t}^a\right)\int_{-\infty}^{\infty}\left( \prod_{i=1}^N \prod_{a=1}^n d
w_i^a\right)\\
&\times&\int_{-i\infty}^{i\infty}\left(\prod_{a=1}^n d\hat\lambda^a\right)  \int_{0}^{\infty} \left(\prod_{t=1}^T\prod_{a=1}^n d
\mu_{t}^a\right) \int_{-\infty}^{\infty}\left(\prod_{t=1}^T \prod_{a=1}^n d\hat{\mu}_{t}^a \right)\exp\left\{\sum_a
\hat\lambda^a(\sum_i w_i^a-N)\right\}
\\
&\times& \prod_{t}\exp\left\{-\frac{1}{2}\sum_{a,b}\hat{\mu}_{t}^aQ_{a,b}\hat{\mu}_{t}^b\right\}
 \exp\left\{\sum_{a,b}\hat{Q}_{a,b}\left(N Q_{a,b}-\sum_i w_i^a w_i^b\right)\right\}\\
&\times& \exp\left\{-\gamma\sum_a(1-\alpha)T \epsilon^a-\gamma\sum_{a,t}
u_{t}^a\right\}\\
&\times&\prod_{t}\exp\left\{i\sum_a\hat{\mu}_{t}^a\left(u_{t}^a+\epsilon^a-\mu_{t}^a \right)\right\}.
\end{eqnarray*}

We can now perform the Gaussian integral over the variables $\{\hat{\mu}_{t}^a\}$:
\begin{eqnarray*}
Z_{\gamma}^n&=&  \int_{-i\infty}^{i\infty}\left(\prod_{a=1}^n\prod_{b=1}^n dQ_{a,b}d\hat Q_{a,b}\right)\int_{-\infty}^{\infty} \left(\prod_{a=1}^n d\epsilon^a\right)
\int_0^{\infty}\left(\prod_{t=1}^T\prod_{a=1}^n du_{t}^a\right)\int_{-\infty}^{\infty}\left( \prod_{i=1}^N \prod_{a=1}^n d
w_i^a\right)\\
&\times&\int_{-i\infty}^{i\infty}\left(\prod_{a=1}^n d\hat\lambda^a\right)  \int_{0}^{\infty} \left(\prod_{t=1}^T\prod_{a=1}^n d
\mu_{t}^a\right) \exp\left\{\sum_a
\hat\lambda^a(\sum_i w_i^a-N)\right\}\\
&\times& \exp\left\{-\gamma\sum_a(1-\alpha)T \epsilon^a-\gamma\sum_{a,t}
u_{t}^a\right\}\exp\left\{\sum_{a,b}\hat{Q}_{a,b}\left(N Q_{a,b}-\sum_i w_i^a w_i^b\right)\right\}\\
&\times& \prod_{t}\exp\left\{-\frac{1}{2}\sum_{a,b}\left(u_{t}^a+\epsilon^a-\mu_{t}^a\right)Q_{a,b}^{-1}\left(u_{t}^b+\epsilon^b-\mu_{t}^b\right)\right\}\exp\left\{-\frac{T}{2}{\rm tr}\log Q\right\}.
\end{eqnarray*}
Introducing the variables $y_{t}^a=\mu_{t}^a-u_{t}^b$ and
$z_{t}^a=\mu_{t}^a+u_{t}^b$ and integrating over the $\{z_{t}^a\}$ we obtain

\begin{eqnarray*}
Z_{\gamma}^n&=&  \int_{-i\infty}^{i\infty}\left(\prod_{a=1}^n\prod_{b=1}^n dQ_{a,b}d\hat Q_{a,b}\right)\int_{-\infty}^{\infty} \left(\prod_{a=1}^n d\epsilon^a\right)
\int_{-\infty}^{\infty}\left( \prod_{i=1}^N \prod_{a=1}^n d
w_i^a\right) \int_{-i\infty}^{i\infty}\left(\prod_{a=1}^n d\hat\lambda^a\right)\\
&\times& \exp\left\{\sum_a
\hat\lambda^a(\sum_i w_i^a-N)\right\} \exp\left\{-\gamma\sum_a(1-\alpha)T \epsilon^a-\gamma\sum_{a,t}
u_{t}^a\right\}\\
&\times&\exp\left\{\sum_{a,b}\hat{Q}_{a,b}\left(N Q_{a,b}-\sum_i w_i^a w_i^b\right)\right\} \prod_{t}\exp\left\{-\frac{1}{2}\sum_{a,b}\left(u_{t}^a+\epsilon^a-\mu_{t}^a\right)Q_{a,b}^{-1}\left(u_{t}^b+\epsilon^b-\mu_{t}^b\right)\right\}\\
 &\times& \exp\left\{-\frac{T}{2}{\rm tr}\log Q-TN\log \gamma +T\log Z_{\gamma}(\{\epsilon^a,Q\})\right\}\end{eqnarray*}
where
\begin{eqnarray*}
Z_{\gamma}(\{\epsilon^a,Q\})&=&\int_{-\infty}^{+\infty} \prod_a dy^a \exp\left\{-\frac{1}{2}\sum_{a,b}(y^a-\epsilon^a)Q_{a,b}^{-1}(y^b-\epsilon^b)\right\}\\
&\times& \exp\left\{\gamma\sum_a y^a\theta(-y^a)\right\}.
\end{eqnarray*}

In order to make further progress, let us consider the replica symmetric ansatz, in accord with the assumption about the uniqueness of the optimum:
\begin{equation}
    Q_{a,b}= \left\{ \begin{array}{cc} q_1 ,&    a =b\\
    q_0 , &  a\neq b \end{array} \right.
\end{equation}
\begin{equation}
    \hat{Q}_{a,b}= \left\{ \begin{array}{cc} r_1 ,&    a =b  \\
    r_0 , &  a\neq b . \end{array} \right.
\end{equation}
Anticipating the behaviour of the various quantities in the limit $\gamma\to\infty$, we introduce the following rescaling:
\bea
 \Delta &=&\gamma(q_1-q_0),\\
 \hat{\Delta}&=&(r_1-r_0)/\gamma,\\
 \lambda^a&=&\hat{\lambda}^a\gamma,\\
 \hat{q}_0&=&r_0\gamma^2.
\eea

The $\vec{w}$-dependent part of the partition function is
\bea
\int [D w] e^{-\gamma F_w} =\int [D w]  e^{\sum_{i a}
\lambda^a w_i^a-\sum_{a,b}\hat{Q}_{a,b}\sum_i w_i^a w_i^b}\qquad.
\eea
Exploiting the identity $\log\langle X^n\rangle\simeq n\langle \log X\rangle$ valid for $n\to 0$,
and after some manipulations, we arrive at the following contribution to the free energy
\be
F_w={\gamma}\Big\langle \log\int\!dw\, e^{-\gamma \left[ \hat{\Delta} w^2-{\lambda} w-z w\sqrt{-2{\hat{q}}_0}\right]}\Big\rangle_z,\qquad
\ee
where the notation $\langle\cdots\rangle_z$ means averaging over the normal variable $z$.
After some further manipulations, we can write the partition function as
\be\label{Zn}
Z_\gamma^n=\int\! d\lambda d \epsilon d q_0 d\Delta d{\hat{q}}_0 d\hat \Delta\, e^{-\gamma n N F[\lambda,{\epsilon},{q}_0,\Delta, {\hat{q}}_0,\hat{\Delta})]}
\ee
where
\bea\label{freeEnergy}
F( \lambda,{\epsilon},{q}_0,\Delta,  {\hat{q}}_0,\hat{\Delta})&=&
\lambda +\tau (1-\alpha)\epsilon -\Delta {\hat{q}}_0-\hat{\Delta} {q}_0\\
\nonumber &-&\frac{1}{\gamma}\Big\langle\log \int_{-\infty}^{\infty} dw e^{-\gamma V(w,z)}\Big\rangle_z
 +\frac{\Delta}{2r\sqrt{\pi}}\int_{-\infty}^{\infty}ds e^{-s^2}
g\left(\frac{\epsilon}{\Delta}+s \sqrt{2 \frac{{q}_0}{\Delta^2}}\right),
\eea

with $r=N/T$,
\be
V(w,z)=\hat{\Delta} w^2-\lambda w -z w\sqrt{-2{\hat{q}}_0}
\ee
 and
\begin{equation}
    g(x)= \left\{ \begin{array}{cc} 0 ,&    x\ge 0\\
    x^2 , &  -1\le x\le 0\\
    -2 x-1, & x<-1
     \end{array} \right..
\end{equation}

The first-order conditions discussed in the main text can be obtained from the above by taking the derivative of $F$ with respect to its arguments (that we call the order parameters) $\lambda,\epsilon,q_0,\Delta,  {\hat{q}}_0$, and $\hat{\Delta}$, and setting these derivatives equal to zero.

\section{Solution of the first order conditions}
\label{sec:appendixB}
\setcounter{equation}{0}

There are some special lines on the $\alpha$ - $r$ plane along which the solution of the first order conditions simplifies significantly. Here, we study four of these lines. (The phase boundary where all three of the order parameters diverge was analyzed in detail in \cite{ciliberti2007On}, so it will not be discussed here.)

\subsection{The $r=0$ axis}
The most obvious case is the interval $0<\alpha<1$ along the horizontal axis. Because $r$=0 is here, we have full information about the underlying process, so we know that all the weights are the same, $w_i=1$, the variance of the weights distribution is zero, hence $\Delta=0$ and $q_0=1$. This is a trivial portfolio and $\epsilon$ is its VaR equal to $\Phi^{-1}(\alpha)$. It is easy to check that these are the solutions of the first order conditions (\ref{equationPhi}-\ref{equationW}) indeed. Moreover, one can calculate the first order corrections in $r$ to them. The results are:
\bea
   q_0 &=& 1 + 2\pi r e^{\left[\Phi^{-1}(\alpha)\right]^2}
              \left(  1-\alpha + \frac1{\sqrt{2\pi}} \Phi^{-1}(\alpha) e^{-\frac12 \left[ \Phi^{-1}(\alpha)\right]^2} \right) \,, \\
  \Delta &=& \sqrt{2\pi} r e^{\frac12  \left( \Phi^{-1}(\alpha)\right)^2} \,, \\
  \epsilon &=& \sqrt{q_0} \Phi^{-1}\left(\alpha-\frac{r}{2}\right) \,. \label{eq:appBr0Epsilon}
\eea
Note that the corrections blow up at both ends of the interval, due to the divergence of  $\Phi^{-1}(\alpha)$ for $\alpha=0, 1$. Eq.~(\ref{eq:appBr0Epsilon}) also requires that $\alpha > r/2$.

\subsection{The $\alpha=1/2$ line}
Let us look for a solution such that $\epsilon = -\Delta/2$. Then, from the symmetries of the functions $\Phi$, $\Psi$ and $W$, Eq.~(\ref{equationSymmetries}),
we immediately see that this will happen along the vertical line $\alpha=0.5$. The two remaining order parameters will be given by
\bea
  \delta \equiv \frac{\Delta}{\sqrt{q_0}} &=& 2 \Phi^{-1}\left(\frac{1+r}{2}\right) \,, \nonumber \\
  \frac1{q_0} &=& \frac{\delta}{\sqrt{2 \pi} r} e^{-\frac12 \left(\frac{\delta}2\right)^2} + \frac{\delta^2}4 - \frac{\delta^2}{2 r}\,.
\eea

\subsection{The $\epsilon=0$ line}
If $\epsilon=0$ the first order conditions become:
\bea
   r &=& \Phi\left(\frac{\Delta}{\sqrt{q_0}}\right) - \frac12 \,, \\
  \alpha &=& \frac{\sqrt{q_0}}{\Delta} \left\{ \Psi\left(\frac{\Delta}{\sqrt{q_0}}\right) - \frac1{\sqrt{2\pi}}  \right\} \,, \\
  1 &=& \Delta^2 \left( 1 - \frac1{2 r} \right) + \frac1{\sqrt{2\pi}} \frac{\Delta \sqrt{q_0}}{r} e^{-\frac{\Delta^2}{2 q_0}} \,.
\eea
We can immediately see that $\Delta/\sqrt{q_0}$ =0 is always a solution. It corresponds to the limit $r=0$ that we have already seen. The non-trivial solution exists for $\alpha> 1/2$, and monotonically increases with $\alpha$, going to $\infty$ as $\alpha$ goes to 1.
The equation of the $\epsilon=0$ line is:
\be
  \alpha = \frac12 +r + \frac1{\sqrt{2\pi}} \frac1{\Phi^{-1}(\frac12+r)} \left[ e^{-\frac12 \left( \Phi^{-1}(\frac12+r)\right)^2}  -1 \right] \,.
\ee
It is shown in Fig.~\ref{fig:epsilon} as the contour line with the parameter 0. It is also shown as the dotted line in the contour map of $\Delta$, Fig.~\ref{fig:Delta}.

\subsection{The $\alpha=1$ line}
There is a further important special line where we can obtain the solution analytically. This is the vertical line at $\alpha=1$. This correponds to the minimax model also called Maximal Loss (ML), the extremal case of ES and a risk measure in its own right \cite{young1998AMinimax}. The key point to notice is that for $\alpha=1$ the order parameter $\Delta$ diverges, but the other two remain finite. This allows us to neglect terms that become exponentially small as $\Delta\to\infty$. Then the first of the first order conditions becomes:
\be
  r = 1 - \Phi\left( \frac{\epsilon}{\sqrt{q_0}} \right)
\ee
that immediately gives
\be
  \frac{\epsilon}{\sqrt{q_0}} = \Phi^{-1}(1-r) = - \Phi^{-1}(r) \,.
\ee
In the following we will use the notation
\be
   \label{eq:appBalp1zeta}
  \rho = - \Phi^{-1}(r)
\ee
and
\be
  h(x) = \frac1{\sqrt{2\pi}} e^{-\frac12 x^2}
\ee
to simplify the formulae.
The second first order condition in the same $\Delta\to\infty$ limit reads
\be
  \alpha \frac{\Delta}{\sqrt{q_0}} = \frac{\Delta}{\sqrt{q_0}} + \rho - \rho \Phi(\rho) - h(\rho)
\ee
yielding
\be
  \frac{\Delta}{\sqrt{q_0}} = \frac{h(\rho)-r\rho}{1-\alpha} \,,
\ee
while the third equation becomes
\be
  \label{eq:appBalph1third}
  \frac1{2\Delta^2} = \frac{\sqrt{q_0}}{\Delta} \frac{\rho}{r} (1-\alpha) + \frac12 \frac{q_0}{\Delta^2} \frac{\rho}{r} \left( r\rho - h(\rho) \right) \,,
\ee
which leads to
\be
  \Delta^2 = r \left[ \frac{2\sqrt{q_0}}{\Delta} \rho (1-\alpha) + \frac{q_0}{\Delta^2} \rho \left( r\rho - h(\rho) \right)  \right]^{-1}  
\ee
Substituting (\ref{eq:appBalph1third}) here gives us the expression for $\Delta$ as
\be
  \label{eq:appBalp1Delta}
  \Delta = \frac1{1-\alpha} \left( \frac{r}{\rho} \left( h(\rho) - r\rho \right)  \right)^{\frac12} \,.
\ee
This shows that $\Delta$, the sensitivity of the portfolio weights to small changes in the returns, becomes infinitely large as we approach the $\alpha=1$ limit. This limit corresponds to the best combination of the worst losses, so to find an infinite sensitivity to returns along this line was to be expected. 

We can now obtain the expressions for the other two order parameters:
\bea
  \label{eq:appBalp1q0}
  \sqrt{q_0} &=& \left( \frac{r}{\rho \left( h(\rho)-r\rho \right)} \right)^{\frac12} \,, \\
  \label{eq:appBalp1eps}
  \epsilon &=& \left(  \frac{r\rho}{h(\rho)-r\rho} \right)^{\frac12} \,.
\eea
The critical behaviour of these quantitites as we approach the point $r=1/2$ may be of interest to record:
\bea
  \Delta &=& \frac1{2\sqrt{\pi}} \frac1{1-\alpha} \frac1{\sqrt{\frac12 - r}} \,,  \\
  \sqrt{q_0} &\approx& \frac1{\sqrt{2}} \frac1{\sqrt{\frac12 - r}} \,, \\
  \epsilon &=& \sqrt{\pi} \sqrt{\frac12 - r} \,.
\eea
As we can see, $\sqrt{q_0}$, the relative estimation error of ES diverges with an exponent -1/2, while $\epsilon$, the VaR of the minimax portfolio, vanishes with an exponent 1/2 in the limit $r\to 1/2$. 

\begin{figure}[h]
  \centering
    \includegraphics[width=45mm]{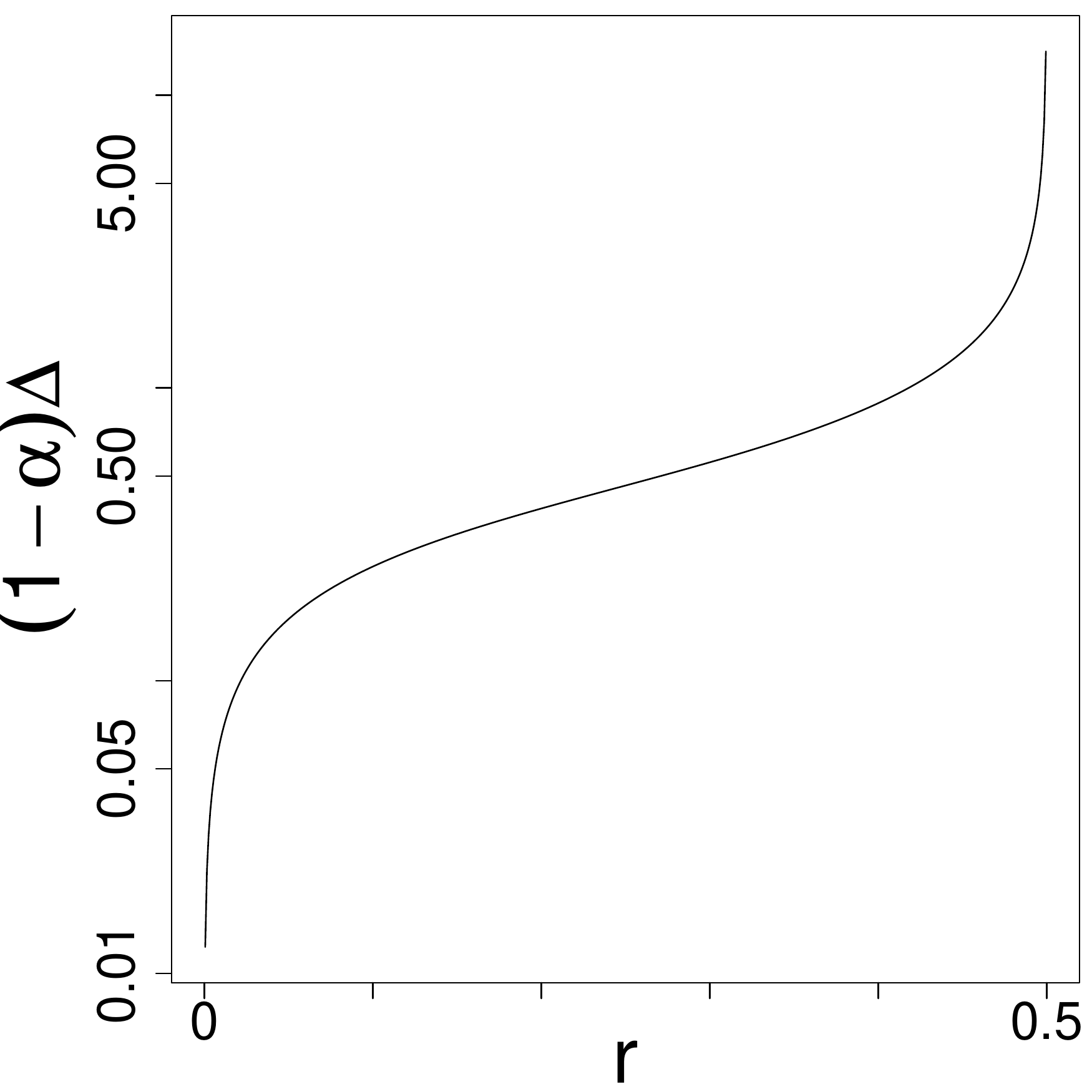}
  \quad
    \includegraphics[width=45mm]{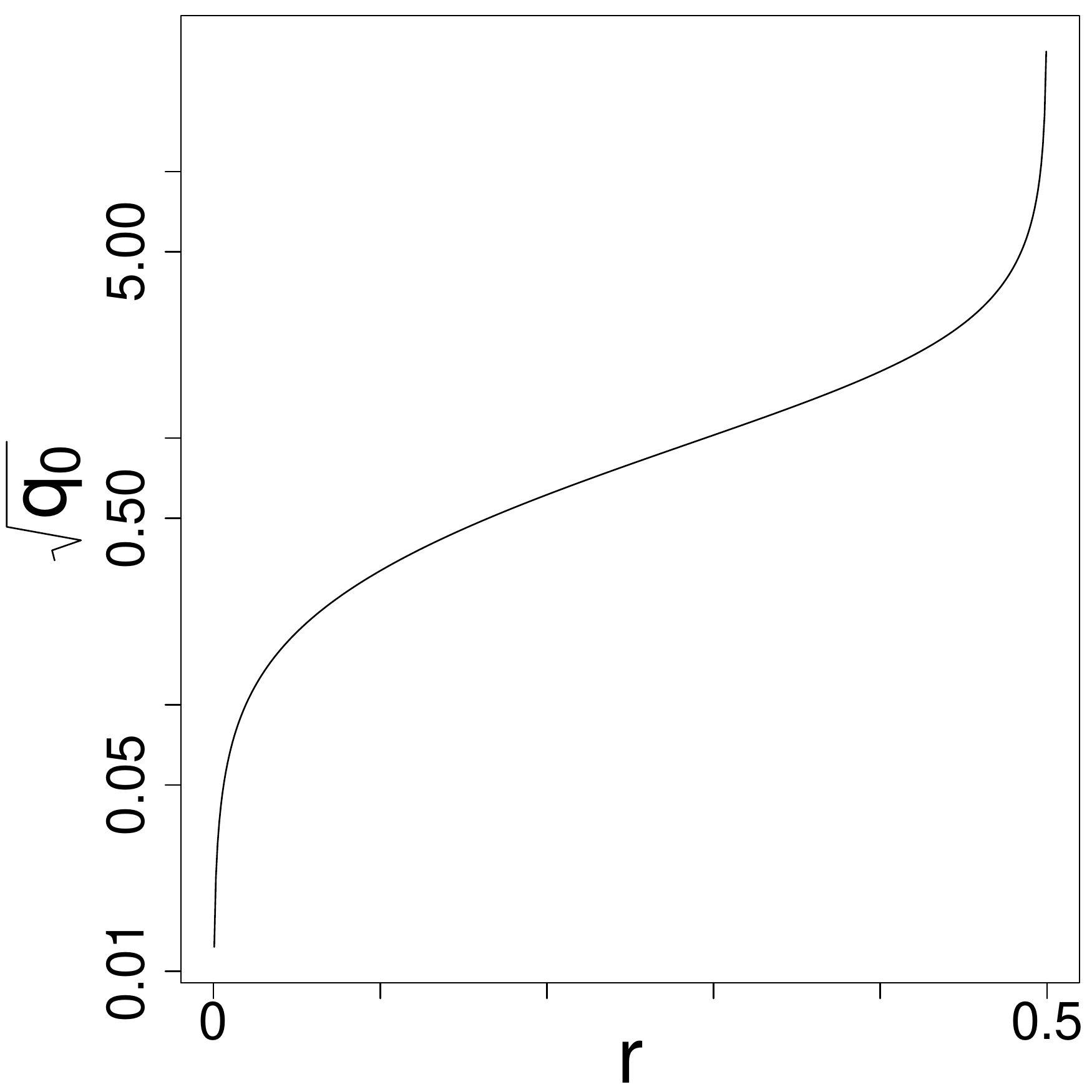}
  \quad
    \includegraphics[width=45mm]{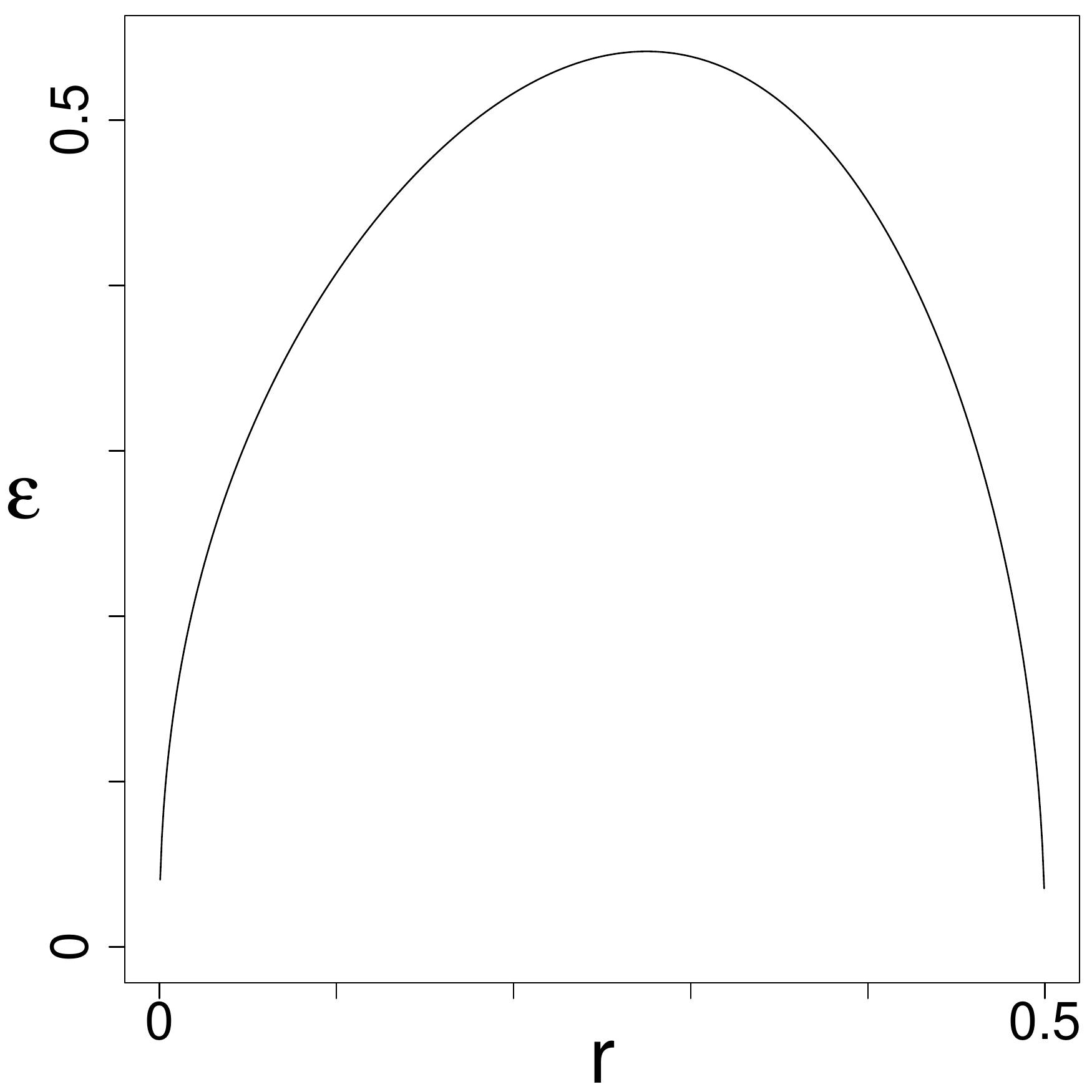}
    \caption{\footnotesize The behaviour of $(1- \alpha)\Delta$ (left), $\sqrt{q_0}$ (center) and $\epsilon$ (right) along the $\alpha=1$ line.}
  \label{fig:alp1vars}
\end{figure}

Eqs.~(\ref{eq:appBalp1Delta})-(\ref{eq:appBalp1eps}) in the $\alpha\to 1$ limit may be rewritten in the following form:
\bea
  (1-\alpha) \Delta &=& \left( \frac{r}{\rho} \left( h(\rho) - r\rho \right)  \right)^{\frac12} \,. \\
  \sqrt{q_0} &=& \frac{r}{\rho} \frac1{(1-\alpha)\Delta} \,, \\
  \epsilon &=& \rho \sqrt{q_0} \,,
\eea
with $\rho$ defined in~(\ref{eq:appBalp1zeta}). The behaviour of these quantities along the $\alpha=1$ line is shown in Fig.~\ref{fig:alp1vars}.

\bibliography{ContourLines}

\end{document}